\documentclass[structabstract]{aa}
\usepackage{graphicx}
\usepackage{epstopdf}
\usepackage{amssymb,amsmath}
\usepackage{aalongtable}
\usepackage{lscape}
\usepackage{color}
\usepackage{natbib}
\usepackage{enumitem}
\usepackage{url}
\usepackage{array}
\newcommand{\gr}{{$\gamma$-ray}}
\newcommand{\lsim}{{\lower.5ex\hbox{$\; \buildrel < \over \sim \;$}}}
\newcommand{\gsim}{{\lower.5ex\hbox{$\; \buildrel > \over \sim \;$}}}
\newcommand{\nupeak}{$\nu_{\rm peak}$}
\newcommand{\nufnupeak}{$\nu_{\rm peak}$f$_{\nu_{\rm peak}}$}

\begin{document}

\title{The 3HSP catalogue of extreme and high-synchrotron peaked blazars\thanks{Tables 1 to 3 are also available in electronic form
at the CDS via anonymous ftp to cdsarc.u-strasbg.fr (130.79.128.5)
or via http://cdsweb.u-strasbg.fr/cgi-bin/qcat?J/A+A/}}


\author{
          Y.-L. Chang  \inst{1,2} \and B. Arsioli \inst{2,3} \and  P. Giommi   \inst{4,5,2}   \and  P. Padovani \inst{6,7} \and C. H.  Brandt \inst{2,8}
         }        
        \institute{Tsung-Dao Lee Institute, Shanghai Jiao Tong University, 800 Dongchuan RD. Minhang District, Shanghai, China
         \and ICRANet, P.zza della Repubblica 10, I-65122, Pescara, Italy
         \and Instituto de F\'isica Gleb Wataghin, Universidade Estadual de Campinas - UNICAMP, Rua S\'ergio Buarque de Holanda 777, 13083-859 Campinas, Brazil 
         \and Italian Space Agency, ASI, via del Politecnico snc, I-00133 Roma, Italy 
         \and Institute for Advanced Study, Technische Universit{\"a}t M{\"u}nchen, Lichtenbergstrasse 2a, D-85748 Garching bei M\"unchen, Germany
         \and European Southern Observatory, Karl-Schwarzschild-Str. 2, D-85748 Garching bei M\"unchen, Germany
         \and Associated to INAF - Osservatorio Astronomico di Roma, via Frascati 33, I-00040 Monteporzio Catone, Italy
         \and Jacobs University, Physics and Earth Sciences, Campus Ring 1, 28759, Bremen, Germany
        \\ \email{ylchang@sjtu.edu.cn}
        }
\abstract
{}
{High-synchrotron peaked blazars (HSPs or HBLs) play a central role in very high-energy (VHE) $\gamma$-ray astronomy, and likely in neutrino astronomy. 
Currently, the largest compilation of HSP blazars, the 2WHSP sample, includes 1691 sources, but it is not complete  in the radio or in the X-ray band. 
In order to provide a larger and more accurate set of HSP blazars that is useful for future statistical studies and to plan for VHE/TeV observations, we present the 3HSP catalogue, the largest sample of extreme and high-synchrotron peaked (EHSP; HSP) blazars and blazar candidates.} 
{We implemented several ways to improve the size and the completeness of the 2WHSP catalogue and reduced the selection biases to be taken into consideration in population studies. 
By discarding the IR constraint and relaxing the radio--IR and IR--X-ray slope criteria, we were able to select more sources with \nupeak\ close to the 10$^{15}$Hz threshold and objects where the host galaxy dominates the flux. 
The selection of extra sources now commences with a cross-matching between radio and X-ray surveys, applying a simple flux ratio cut. 
We also considered \textit{Fermi}-LAT catalogues to find reasonable HSP-candidates that are detected in the $\gamma$-ray band but are not included in X-ray or radio source catalogues. 
The new method, and the use of newly available multi-frequency data, allowed us to add 395 sources to the sample, to remove 73 2WHSP sources that were previously flagged as uncertain and could not be confirmed as genuine HSP blazars, and to update parameters obtained by fitting the synchrotron component.}
{The 3HSP catalogue includes 2013 sources, 88\% of which with a redshift estimation, a much higher percentage than in any other list of HSP blazars. 
All new \gr\ detections are described in the First and Second Brazil ICRANet $\gamma$-ray blazar catalogues (1BIGB \& 2BIGB) also taking into account the 4FGL list of $\gamma$-ray sources published by the \textit{Fermi} Large Area Telescope (\textit{Fermi}-LAT) team.
Moreover, the cross-matching between the 2WHSP, 2FHL HSP, and IceCube neutrino positions  suggests that HSPs are likely counterparts of neutrino events, which implies the 3HSP catalogue is also useful   in that respect. 
The 3HSP catalogue shows improved completeness compared to its predecessors, the 1WHSP and 2WHSP catalogues, and follows the track of their increasing relevance for VHE astronomy.}  
{}

 
\maketitle
%

\section{Introduction}
\label{intro}
Blazars are a class of active galactic nuclei (AGN) characterised by rapid and large amplitude spectral variability, assumed to be due to the presence of a relativistic jet pointing very close to the line of sight \citep{Blandford1978, Antonucci1993, Urry1995}. 
The emission of these objects is non-thermal over most of the electromagnetic spectrum, from radio frequencies to hard $\gamma$-rays. 
The observed radiation shows extreme properties, mostly coming from relativistic amplification effects. 
The observed spectral energy distribution (SED) has a typical shape composed of two broad humps, one peaking between the far-infrared  and the soft X-ray band, due to synchrotron emission, and the other peaking in the hard X-ray to \gr~bands. 

If the peak frequency of the synchrotron component (\nupeak) in $\nu-\nu{\rm f}_{\nu}$ space is higher than $10^{15}$~Hz, a blazar is usually called high-synchrotron peaked blazar (HBL or HSP) \citep{Padovani1995,Abdo2010}
;in particular, an HSP blazar with a synchrotron peak frequency above $10^17$ Hz is known as extreme high-energy synchrotron peaked (EHSP) \citep[EHSP;][]{Giommi1999,Costamante2001}. 
Although dozens of objects clearly peaking at frequencies as high as $10^{18.5}$\,Hz ($\sim$5-10\,keV) have been found, evidence for the synchrotron peak reaching the MeV range is still under debate \citep{Tanaka2014,Kaufmann2011,Tavecchio2011,Giommi2001}. 

Observations have shown that both EHSPs and HSPs are bright and extremely variable sources of high-energy \gr\ and TeV photons (TeVCat\footnote{\url{http://tevcat.uchicago.edu}}) and that they may be the dominant component of a putative extragalactic TeV background \citep{Padovani1993,Giommi2006,DiMauro2014,Giommi2015,Ajello2015}. 
Given that most of the extragalactic objects detected so far above a few TeV are HSPs \citep[][see also TeVCat]{Giommi2009,Padovani2015,Arsioli2015,Fermi3FHL2017}, HSP blazars are the main targets for future \gr\ and VHE observations \citep[see examples in][]{Chang2017,Arsioli2017}. 


\citet{Arsioli2015} (Paper I) built a catalogue of HSP blazars  based on data from the Wide-field Infrared Survey Explorer mission (WISE); it was called  1 WISE HSP (1WHSP) and only selected sources  inside a specific area of the colour-colour diagram called {\it Sedentary WISE colour domain} (SWCD), which was defined as an extension of the WISE blazar strip \citep{Massaro2011,DAbrusco2012,Massaro2012} in order to include all sources from the Sedentary survey \citep{Giommi1999,Giommi2005,Piranomonte2007}. 
They cross-matched the  sources in the AllWISE catalogue \citep{Cutri2013} that are inside the SWCD with different radio and X-ray catalogues using TOPCAT\footnote{\url{http://www.star.bris.ac.uk/~mbt/topcat/}}, applied three spectral slope criteria, and selected sources with \nupeak~$> 10^{15}$~Hz and Galactic latitude $b>|20^\circ|$. 
The slope criteria applied in Paper I are the radio to IR spectral slope, the IR to X-ray spectral slope, and the AllWISE channels W1 to W3 spectral slope\footnote{$0.05<\alpha_{1.4{\rm GHz}-3.4\mu{\rm m}}<0.45,0.4<\alpha_{4.6\mu{\rm m}-1{\rm keV}}<1.1, \\{\rm and} -1.0< \alpha_{3.4 \mu{\rm m}-12.0\mu{\rm m}}<0.7$}; the criteria are obtained from normalised and rescaled SEDs of three well-known HSP blazars. 

 \citet{Chang2017} (Paper II) assembled the 2WHSP catalogue, an updated version of  1WHSP  extended to lower Galactic latitudes ($b>|10^\circ|$) and including bright HSPs  in the region close to the Galactic plane. 
Similarly to Paper I, the 2WHSP catalogue was constructed by cross-matching three radio catalogues \citep[NVSS, FIRST, and SUMSS:][]{Condon1998,White1997,Manch2003} with the AllWISE IR catalogue and then with various X-ray catalogues \citep[RASS BSC and FSC, 1SWXRT and deep XRT GRB, 3XMM, XMM slew, Einstein IPC, IPC slew, WGACAT, Chandra, and BMW:][]{Voges1999,Voges2000,DElia2013,Puccetti2011,Rosen2015,Saxton2008,Harris1993,Elvis1992,White2000,Evans2010,Panzera2003}. 
However, the 2WHSP catalogue does not apply the WISE colour-colour diagram, and any IR slope criteria (e.g. the W1-W3 slope as measured from AllWise). 
This was done to avoid missing several HSPs where the IR and optical bands are dominated by the host galaxy thermal radiation, as many of the blazars that are classified as 5BZG in the latest BZcat catalogue \cite{Massaro2015}. 

\citet{Chang2017} used the SSDC SED tool\footnote{\url{http://tools.ssdc.asi.it/SED}} to examine and fit the synchrotron component with a third-degree polynomial to get \nupeak~and synchrotron peak flux (\nufnupeak) values for each candidate. 
The 2WHSP  catalogue includes a total of 1,691 sources with 540 previously known HSPs, 288 new HSPs, and 814 HSP-candidates. 
The name WHSP, which stands for WISE high-synchrotron peaked blazars, indicates that  all sources in 2WHSP (except for one, 2WHSP J135340.2$-$663958.0) have WISE counterparts. 
For each 2WHSP source, we adopted as the best coordinates those taken from the WISE catalogue. 

The 2WHSP catalogue has been applied as a seed to HE and VHE observations to find new VHE detections or counterparts of VHE catalogues. 
\citet{Arsioli2017} analysed bright 2WHSP sources using archival {\it Fermi}-LAT Pass 8 data integrated over 7.2-year observations. 
By using the positions of 2WHSP sources as seeds for the data analysis, they found 150 new \gr\ detections not yet reported in any of previous $\gamma$-ray catalogues. 
The 150 new detections are collected in the  First Brazil ICRANet gamma-ray blazar catalogue (1BIGB). 

Moreover, \citet{Padovani2016} cross-matched the 2WHSP and the HSP subsample of the  second catalogue of Hard {\it Fermi}-LAT Sources (2FHL)  \citep{Ackermann2015} with IceCube neutrinos \citep{IceCube2015} suggesting that, among the blazar family, HSPs are the most probable counterparts for astrophysical neutrinos. 
\citet{Resconi2017} have presented further evidence of a connection between 2FHL HSPs, with very high-energy neutrinos and ultra high-energy cosmic rays (UHECRs) when cross-matching the  2FHL HBL  subsample with UHECRs from the Pierre Auger Observatory \citep{Abraham2004,Pierre2015} and the Telescope Array \citep{Abu-zayyad2012}. 
In a nutshell, HSP catalogues are important for HE, VHE, and multi-messenger astronomy. 
Their statistical properties, such as completeness, evolution, and possible bias associated with the building of HSP catalogues need to be analysed carefully. 

Throughout the paper we adopt a Flat-$\Lambda$CDM cosmology with the following parameters \citep{Carroll1992}: $\Omega_{\rm m}=0.3~{\rm and}~H_{0}=70~{\rm km}~{\rm s}^{-1}~{\rm Mpc}^{-1}$. Poisson errors  of
$1\sigma$  \citep{Gehrels1986} were applied when the numbers were $\le50$ in the $\log$N-$\log$S. 

\section{Missing sources in the 2WHSP catalogue}

\begin{figure} [h!]
\centering
\includegraphics[width=0.98\linewidth]{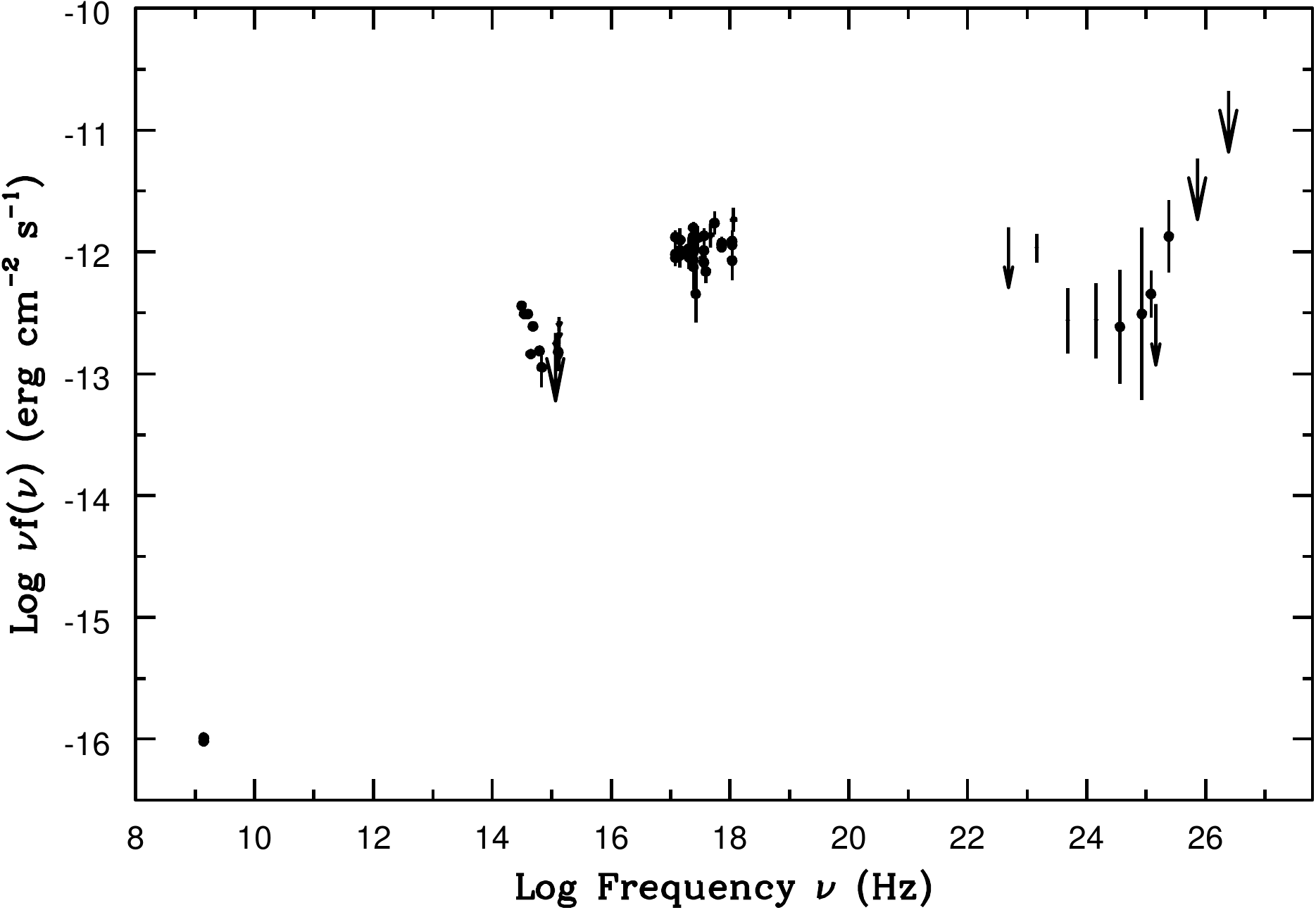}\\
\includegraphics[width=0.98\linewidth]{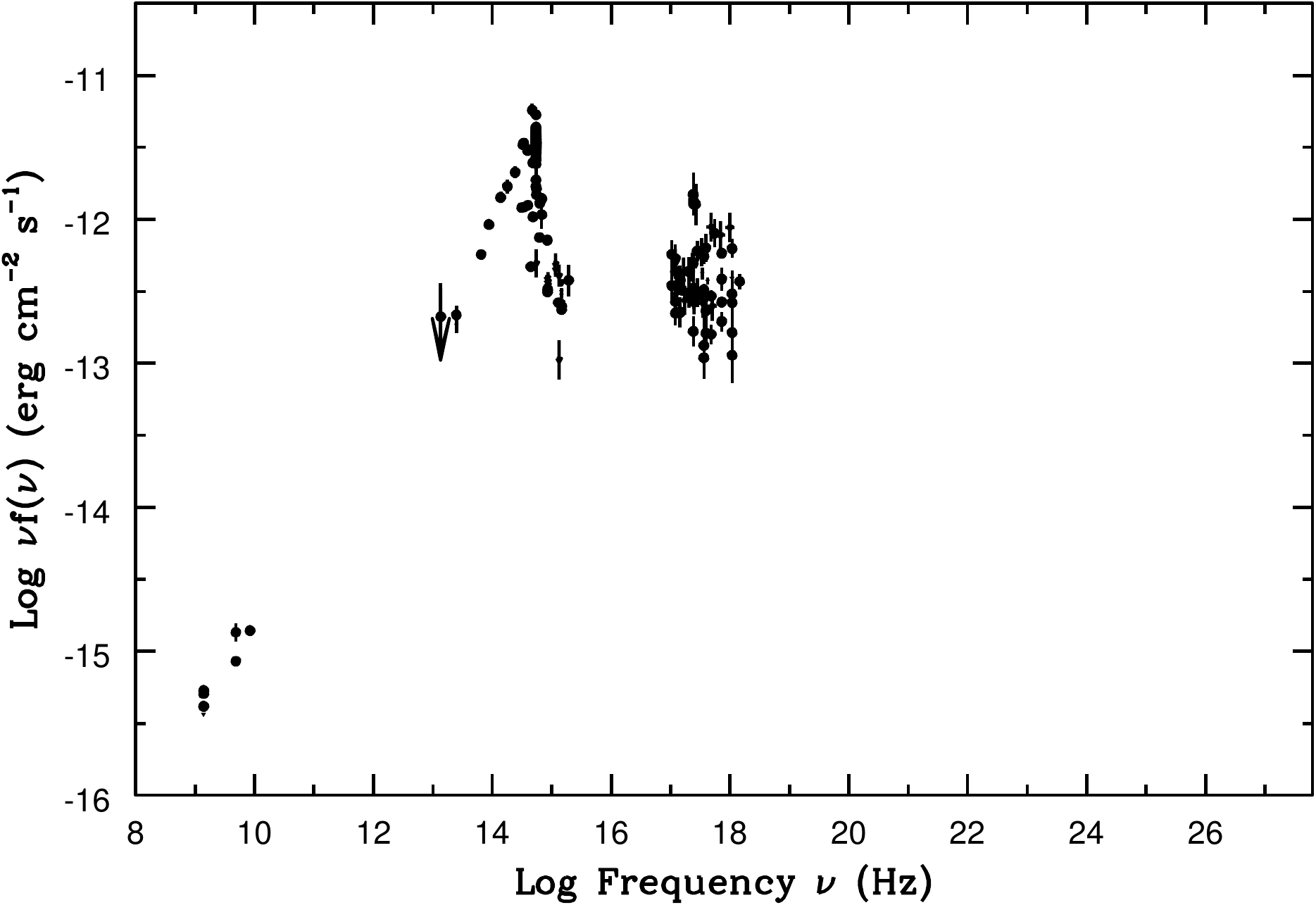}
\caption[The SEDs of 5BZB J0403-2429 and 5BZG J0903+4055]{Spectral energy
distributions of 5BZB J0403-2429 (top) and 5BZG J0903+4055 (bottom). The  sources were not selected for 2WHSP  due to  the absence of IR data (top) and  the slope criterion (bottom).}
\label{bzcatmiss}
\end{figure}

There are a number of known HSP blazars that are not included in the 2WHSP catalogue, in particular all those that have not been detected in WISE surveys. 
For example, the blazar 5BZB J0403-2429 (Fig. \ref{bzcatmiss}, top) \citep{Massaro2015} does not have an IR counterpart but is an extreme HSP with \gr\ counterparts in {\it Fermi} Third Catalog of Hard Fermi-LAT Sources (3FHL)\citep[3FHL J0403.2$-$2428:][]{Fermi3FHL2017} and {\it Fermi} LAT 8-year Source Catalog (4FGL\footnote{\url{https://fermi.gsfc.nasa.gov/ssc/data/access/lat/8yr_catalog/}}) catalogues. 
After checking all the HSP-candidates without IR detection from WISE, we realised that all of them were relatively close to another bright source and it was  likely the  reason why they were not included in the WISE source list. 
However, all of them have optical counterparts from the Panoramic Survey Telescope and Rapid Response System (PANSTARRs) or The United States Naval Observatory (USNO) catalogue, and may be robust HSP-candidates or have been catalogued as blazars, like 5BZB J0403-2429.
This suggest that the presence of a nearby bright IR source affects the IR detections around it, and therefore offers a good explanation for the reason why several promising HSP-candidates have no WISE counterpart.

\begin{figure} [h!]
\centering
\includegraphics[width=0.98\linewidth]{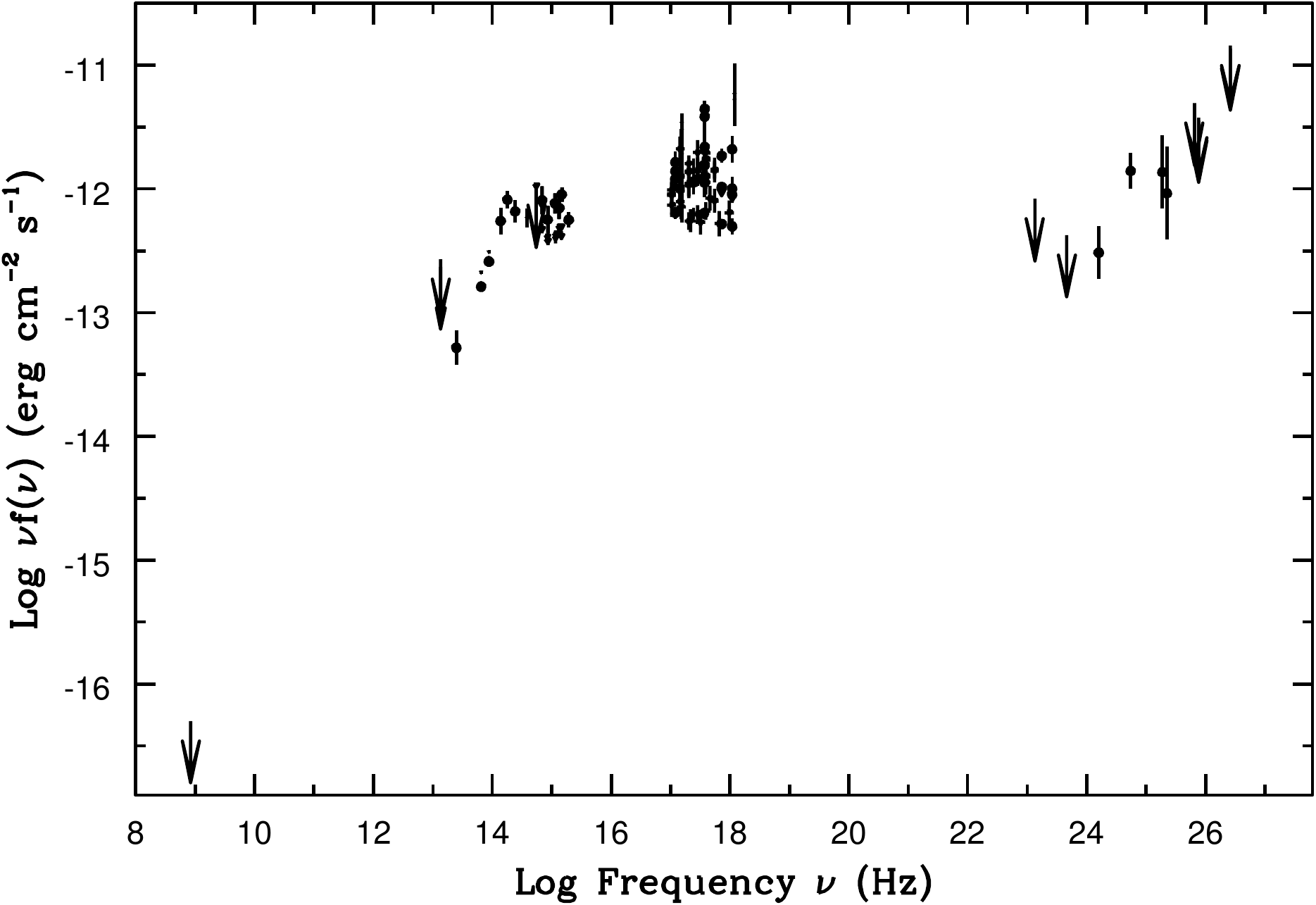}
\caption[The SED of 6dF J0213586-695137]{Spectral energy
distribution of 6dF J0213586-695137. This source was missed by 2WHSP because there is no radio data available for the source.}
\label{norrmiss}
\end{figure}

Moreover, sources where a bright host galaxy dominates the IR and optical bands have low $\alpha_{\rm 1.4 GHz-3.6\mu m}$ values and therefore do not satisfy the radio-IR slope criterion adopted for the selection of 2WHSP blazars. 
This applies for instance to the case of 5BZG J0903+4055 (Fig. \ref{bzcatmiss}, lower panel), which is not in the 2WHSP catalogue since it does not satisfy the slope criterion ($0.05<\alpha_{1.4{\rm GHz}-3.4\mu{\rm m}}<0.45$) owing to host galaxy contamination in the IR band, even thought it is  an HSP with a highly variable X-ray spectrum and a luminosity that ranges between 2.4 and 7.3$\times 10^{44} \rm{erg}~\rm{s}^{-1}$ in the 0.3-10.0 keV band \citep{giommi2019}, reflecting changes in \nupeak. 
 
Finally, some sources without radio data could be suitable HSP-candidates with extreme \nupeak\ values. 
For example, the source 6dF J0213586-695137 (Fig.~\ref{norrmiss}) is clearly an HSP with strong X-ray $\gamma$-ray  emission, but with no radio counterpart in the Sydney University Molonglo Sky Survey (SUMSS) catalogue. 
The {\it Fermi} Pass 8 analysis (triangle points), 3FHL, and 4FGL data show that the $\gamma$-ray fluxes are consistent with data from the lower energy bands, suggesting that its inverse Compton emission might peak in the TeV band. 
This type of very faint radio source clearly could not be selected when using the criteria adopted for 2WHSP.

\section{Building a more complete catalogue}
\label{sec:building}


As discussed above, the 2WHSP catalogue still misses some relevant HSPs.
Among the factors that cause this deficiency, the slope criterion is the most relevant; also, as Fig. \ref{bzcatmiss} shows, an HSP is not necessarily  detectable by WISE. 
To recover these missing sources in this work we do not demand that all candidates  have an IR counterpart; instead,  we  only apply the X-ray--to--radio flux ratio criterion. 
We note that for the new catalogue, all criteria are applied to find more blazars and not to eliminate sources that were already in previous WHSP catalogues. This was done only when we found new data implying that the sources are not HSPs.

\subsection{Cross-matching and radio and X-ray slope criteria}
\label{crossmatch}
The first step we  took to increase the 2WHSP completeness was to cross-match the RASS and NVSS catalogues using a fixed 0.8 arcmin radius (which is larger than the positional uncertainty of more than 99\% of the RASS sources) and to choose those matches for which the X-ray to radio flux ratio $\rm{f_{\rm x}/f_{\rm r}}>9\times10^{-11} \rm{erg}~\rm{cm}^{-2}~\rm{s}^{-1}~\rm{Jy}^{-1}$. 
The X-ray RASS flux used in this work were  corrected for Galactic absorption.
We note  that this value is obtained from the average radio--to--X-ray flux ratio of 2WHSP sources with \nupeak~ close to $10^{15}$~Hz. 
Sources that are already part of the 2WHSP catalogue and those that are close to the Galactic plane ($|b| \le 10^\circ$) were excluded. 
This procedure led to a list of 3011 additional pre-selected candidates.

We then cross-matched this list with the 5BZCat \citep{Massaro2015}, the  XMM-Newton Optical Monitor Serendipitous UV Source Survey   \citep[XMMOM][]{Page2012}, and the {\it Fermi} 3FHL  \citep{Fermi3FHL2017} catalogues, using radii of 0.3, 0.3, and 20 arcmin, respectively (i.e.  values that are somewhat larger than the largest positional uncertainties of these catalogues) to identify HSPs that have already been catalogued, and to check on all candidates having UV or \gr\ non-thermal emission. 
We used XMMOM and {\it Fermi} data to reduce the pre-selected sample since a source with UV or $\gamma$-ray detection is more likely to be a blazar.
These cross-matches were not performed to exclude blazars that were already listed in other catalogues, on the contrary, they were done to help us find more HSP blazars, especially those already in 5BZCat or with a $\gamma$-ray (Fermi 3FHL) or UV (XMMOM) counterpart. 
This process led to 254 pre-selected sources, of which  30 are in 5BZCat, 149  in XMMOM, and 75  in Fermi 3FHL. 
Ten sources are in both 5BZCat and 3FHL, while three sources are in both XMMOM and 3FHL. 

We note that all the cross-matchings described above use fixed radii that are relatively large compared to the average positional error of each catalogue. This was done to avoid loosing candidates in the first step of the selection process. Further refinement of all the tentative associations is carried out through the examination of the error circle map of each candidate.
This map is produced by the VOU-Blazar tool (available at the Open Universe website\footnote{\url{http://openuniverse.asi.it}}
and on GitHub\footnote{\url{https://github.com/ecylchang/VOU_Blazars}}), which generates a spatial-plot where all the sources in the field and their position uncertainties (error regions) are visualised.
After thoroughly checking this map for error circle consistencies, we inspected the SEDs of all the pre-selected candidates before making a final decision. 
Along with the SED, we also checked the optical spectrum, the bibliography on NED, and the radio, IR, and optical images for every source in order to classify each of them. 
Moreover, we excluded sources close to clusters of galaxies or members of a galaxy cluster, and removed all those with extended X-ray emission. 
For this purpose we cross-matched our list with a number of catalogues of clusters of galaxies \citep[Zwicky, PLANCK, and MCXC: ][]{Zwicky1968,PlanckCollaboration2016b,Piffaretti2011}, and carefully checked the X-ray image for every 3HSP source, when available (see section 2.4 in Paper II).
Overall, we expect to have very few spurious sources due to contamination of X-rays from the hot gas in clusters. 
After the  inspection of the 254 pre-selected sources, 58 of them were determined to be classifiable as HSPs or HSP-candidates and  were added to the current HSP catalogue. 

There are still 2757 radio--X-ray matching sources to be checked.
In this version of the catalogue we added the 58 sources that have a \gr~detection or that have  already been listed in 5BZCat. 
Given that nearly half of the sources in our HSP catalogue have been detected in the \gr\ band, we expect that there should be approximately 60 HSPs with \gr\ flux below the Fermi-LAT sensitivity among the other 2,757 sources.
In a future update of the catalogue we will use the VOU-Blazars tool (see section~\ref{voublazar}) to carefully examine  the remaining 2,757 radio--X-ray matching sources. 


\subsection{Searching for extra sources using {\it Fermi} \gr\ catalogues}
We note that robust HSP-candidates, especially those with \gr\ detection, do not necessarily require both radio and X-ray data to be present in the currently available archives, as shown by Figure~\ref{norrmiss}.
 There are still several sources that cannot be selected on the basis of the radio/X-ray flux ratio, but that are detected in \gr s.
To identify these HSPs, a careful examination of all the sources in the 3FHL catalogue and  their possible HSP counterparts was performed. 
The recently released Fermi 4FGL catalogue\footnote{\url{https://fermi.gsfc.nasa.gov/ssc/data/access/lat/8yr_catalog/}} \citep{4FGL} was also searched in an effort to find new HSP sources.
HSPs typically have \gr\ photon index $\Gamma < 2.0$ \citep[see e.g. the {\it Fermi}-3LAC catalogue;][]{Ackermann2015a}, thus we checked those 4FGL \gr\ sources with hard \gr\ slopes. 

To search for additional HSPs among {\it Fermi} sources, we assumed that a Fermi $\gamma$-ray detected blazar has a counterpart in existing radio or X-ray catalogues, but not necessarily in both.
 Thus, we checked every radio and X-ray source around the {\it Fermi} detections with the inspection tools available via the Open Universe portal. 
Then we searched for possible optical, IR, and UV counterparts for them. 
Even though there might be more than one optical source within the Fermi error region, we   inspected those that have a radio or X-ray counterpart.
After that we carefully examined every possible counterpart with a multi-frequency approach (see section~\ref{crossmatch}). 
This lead to a total of 168 HSPs that were not in the 2WHSP sample, but were in 3FHL. 
We note that here the number is approximately three times larger than in the last step (168 versus 58 sources), suggesting that most of the sources with radio and X-ray matches had  already been selected in previous steps. 
Moreover, out of the 389 4FGL sources with a hard spectrum that are still out of our selection, 121 HSP blazars or candidate blazars were identified and added to our catalogue after carefully checking their SEDs. 

As expected, approximately half of the sources added   from the 3FHL or the 4FGL catalogue do not have an associated radio or X-ray detection.
These sources are either without radio or X-ray data or have an X-ray--to--radio flux ratio that is not typical of previously known HSPs (i.e.  ${\rm f_x/f_r} <9\times10^{-11}$). 
We  discuss the selection of sources without a radio counterpart in the Appendix.
In conclusion, we only select HSP-candidates after a careful review of each field, using all available information from multi-frequency catalogues.

\subsection{The VOU-Blazar tool}
\label{voublazar}
To increase the efficiency of our search for new HSPs, we developed a tool called VOU-Blazars \footnote{\url{https://github.com/ecylchang/VOU_Blazars}} that uses Virtual Observatory (VO) protocols to retrieve multi-frequency information from a large number of services distributed worldwide and combines the retrieved data to find sources with SEDs similar to those of blazars.
This tool allowed us to select 48 additional HSP-candidates. 
Most of these new sources have been detected only in recent  {\it Swift} X-ray telescope (XRT) or X-ray Multi-Mirror Mission ({\it XMM-Newton}) observations, and therefore are not included in the catalogues used by the SSDC-SED builder tool provided by the Italian Space Agency's  Space Science Data Center (ASI-SSDC).
With the VOU-Blazars, it is also  possible to retrieve  results from recent X-ray observations, and this allowed us to significantly improve our search with respect to the 2WHSP catalogue. 
The VOU-Blazars tool has already been used to
locate all the probable and confirmed blazars within the uncertainly region of the astrophysical neutrino IceCube-170922A \citep{Padovani2018}.

\subsection{SED selection criteria and the estimation of synchrotron peak}
The process discussed above led to the addition of a total of 395 
new sources to the updated HSP catalogue.
Only objects having non-thermal data in at least three different bands were considered. 

We determine the \nupeak\ for sources without a radio or X-ray counterpart as follows: 
for sources without an X-ray measurement we verified that 
\nupeak\ was higher than $10^{15}$Hz using radio, IR, and UV data. 
For sources without a counterpart in current radio catalogues we used the (non-thermal) IR, UV, and X-ray data (see the Appendix for more details). 
If we were able to identify a non-thermal component in the IR, UV, or  X-ray bands, we estimated the \nupeak~value by fitting those data points. 
This was done only when the source had at least one data point in the X-ray band, otherwise a lower limit or an uncertain value was derived. 
Only a very few candidates without radio data were  found; in these rare cases we used the UV and X-ray data to estimate the synchrotron peak position. 
In cases where no UV data were available, the synchrotron peak was estimated only if a good quality X-ray spectrum (producing several SED good signal-to-noise points) could be found in the {\it Swift XRT} or {\it XMM-Newton} public archives.

The criteria used for source selection are somewhat inhomogeneous given that different data are available for each source. 
In practice, we selected a source based on the following criteria: 
\begin{itemize}
\item the SED includes at least three data points that can be attributed to non-thermal emission from, for example, radio, IR/optical, X-ray;  radio, optical/UV, X-ray;  IR, optical/UV, X-ray; or radio, IR, UV;
\item the optical spectrum and the SED is blazar-like (see e.g. Fig. 1 of \cite{AGNReview}) and the non-thermal data are sufficient to reliably establish that \nupeak~$\ge 10^{15}$; 
\item the source is included in a blazar catalogue such as 5BZCat or Sedentary; 
\item the radio-to-synchrotron peak flux ratio is similar to that of standard HSPs, i.e. $\approx 4-5$ orders of magnitude; 
\item the radio emission is within the X-ray position error, and the optical, IR, UV counterparts coincide with the radio emission; 
\item the source was detected in a very high-energy survey and the SED built with the \gr\, and the lower energy data is consistent with that of an HSP blazar. 
\end{itemize}

\subsection{Estimation of spurious sources}
\label{spurious}
 The contamination of the 3HSP sample with other types of objects is very much dependent on the position uncertainties of each individual multi-frequency detection. 
For the \gr\ detected sources, the number of expected spurious associations is less than three, while for other sources it depends on the quality of the available X-ray data, which is   heterogeneous.
The uncertainties are mostly dominated by X-ray sources that have only been detected in the RASS survey, given that for those cases the positional error can be relatively large, up to $\sim 40$ arcseconds. 
For sources with a more precise positioning (1-5 arcseconds), like those detected by {\it XMM-Newton} or {\it Swift}, the expected number of spurious sources is very small, likely less than one. 

The number of spurious associations between 3HSP objects and \gr\ sources can be estimated from the {\it Fermi} positional uncertainty, the number of {\it Fermi} sources, and the number density of HSPs. 
For the 4FGL sources above the Galactic plane ($b\ge|10|^\circ$), the average error ellipse major axis is 4.46 arcmin, and the average error ellipse minor axis is 3.59 arcmin. 
Thus, the average area associated with the positional uncertainty of 4FGL sources at $b\ge|10|^\circ$ is $\sim 0.014~{\rm deg}^{2}$. 
There are 3663 4FGL sources with $b\ge|10|^\circ$, so the total area covered by 4FGL error regions is $0.014 \times 3663 = 51.3~{\rm deg}^{2}$.
The radio logN-logS of 3HSP gives a surface density of 3HSP objects with radio flux $> 3.5$~mJy of $\approx~0.05~{\rm deg}^{-2}$. 
Multiplying the total area $51.3~{\rm deg}^{2}$ by the number density 0.05, we get 2.6.
This should be seen as an upper limit since this calculation assumes that all 3HSP sources have a radio flux density of 3.5 mJy. 
Obviously, the radio flux density of the large majority of our sources is significantly higher than the limit of current radio surveys. 
In conclusion, the expected number of spurious association of 3HSP objects with (high b) Fermi 4FGL sources is less than 2.6 sources, that is $< 2.6/2011$ or $<  0.15\%$. 

The expected number of spurious associations related to the  radio--X-ray matching sources can be estimated as follows.
The average uncertainty radius in the RASS survey is 18.8 arcsec, corresponding to an area of 8.6$\times 10^{-5}{\rm deg}^{2}$. 
This value, multiplied by the number of sources in the RASS catalogue at $b\ge|10|^\circ$ and Dec $> -40^\circ$, gives an area of $7.6~{\rm deg}^{2}$, corresponding to the surface covered by all RASS error regions in the part of the sky covered by the NVSS survey.
Given that in the NVSS catalogue at $b\ge|10|^\circ$ there are $\sim43.5$ sources per square degree, the expected number of random NVSS/RASS matches is $\sim$ 330. 
Since only 64\% of sources in the NVSS catalogue matches a WISE source, and only 14.5\% of the WISE sources are located in the SWCD area, the expected fraction of NVSS/RASS matches including a WISE source in the SWCD area is 0.09. 
The total number of expected spurious associations that would meet our selection criteria is therefore 30.6 = 330$\times$0.09, equivalent to less than 2\% of the 3HSP sources at Dec $>-40^\circ$.


The equivalent calculation applied to the  subsample of 3HSP objects with no infrared  counterpart but with UV information, with WISE replaced by Galaxy Evolution Explorer (GALEX) and the SWCD replaced by UV to X-ray spectral slope constraint, gives an expected value of about nine spurious sources.
These numbers should be considered conservative upper limits as in most cases, in addition to the basic multi-frequency data considered above, we have  information such as radio measurements at different frequencies, X-ray data with much better positioning than that of the RASS survey and \gr\ data, for example, that is fully consistent with the assumption of  an HSP blazar, both in terms of positional uncertainties and SED constraints.
All sources with poor multi-frequency coverage and large X-ray positional uncertainty have been flagged as `candidate'.

A careful inspection, based on newly available multi-frequency data, has been carried out on the sources of the 2WHSP catalogue, resulting in the elimination of 73 objects (4.32\% of the total) as spurious associations or blazars with intermediate \nupeak\ energy. 
As explained above we expect the fraction of incorrect 3HSP associations to be significantly lower than that of the 2WHSP catalogue.

\subsection{Comparison between new sources and the 2WHSP catalogue}

\begin{figure} [h!]
\centering
\includegraphics[width=0.9\linewidth]{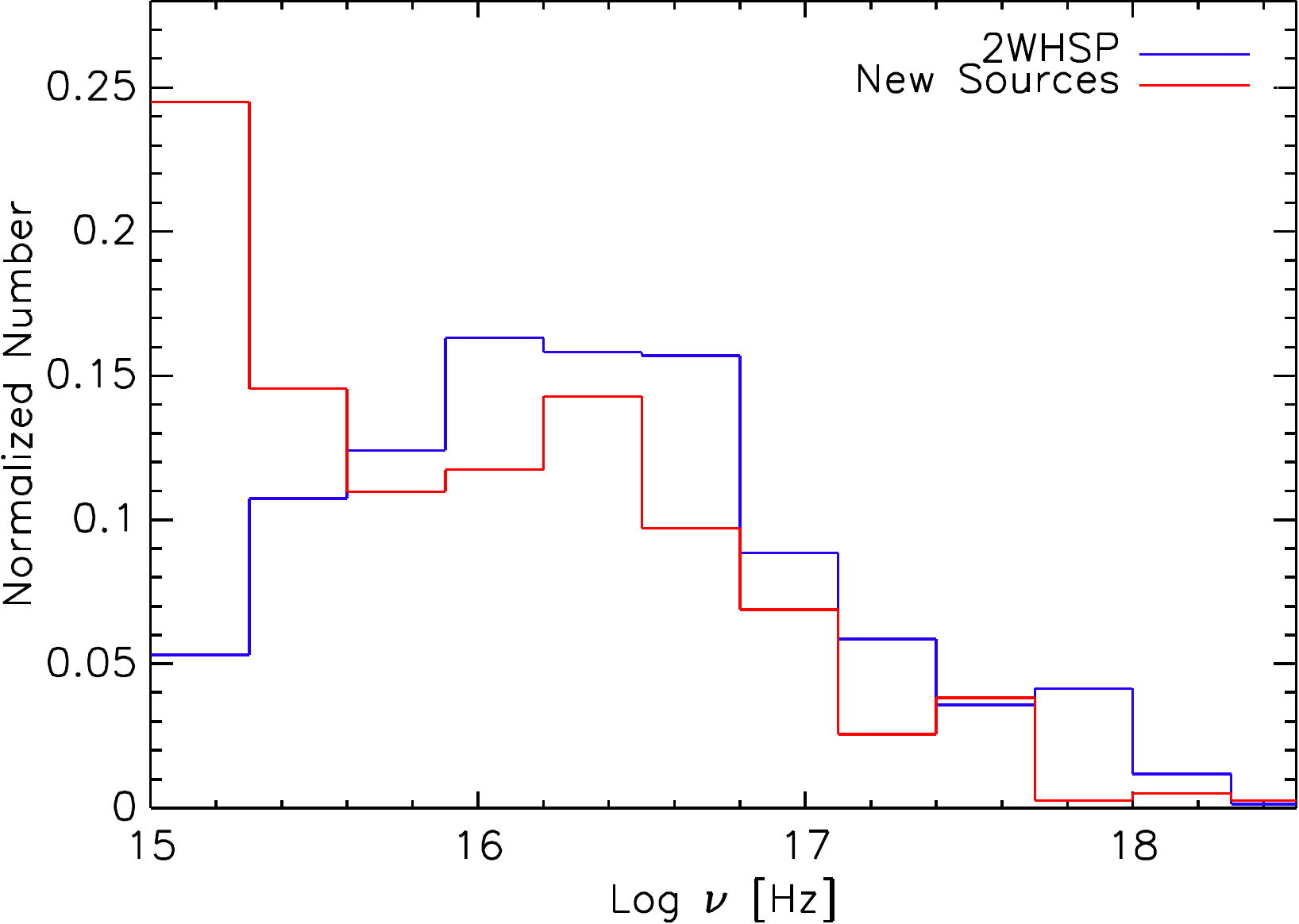}
\includegraphics[width=0.9\linewidth]{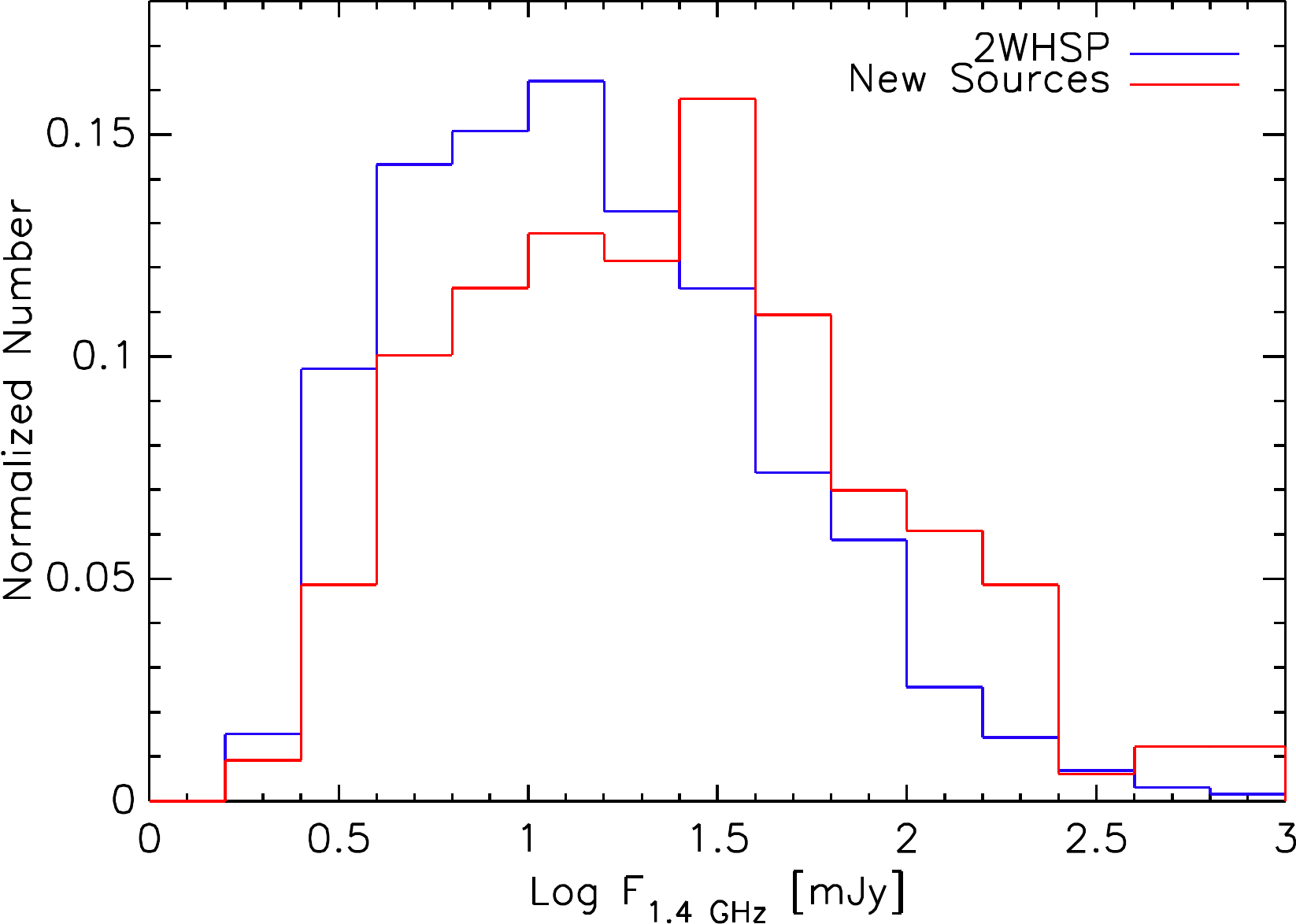}
\caption[The \nupeak\ and radio flux density distribution of the new sources and those of the 2WHSP catalogue]{Distribution of the new sources and those in the 2WHSP catalogue (\nupeak,\ top;  radio flux density,
bottom) normalised according to the number of objects in each sample.}
\label{diffnumjy}
\end{figure}

\begin{figure} [h!]
\centering
\includegraphics[width=0.9\linewidth]{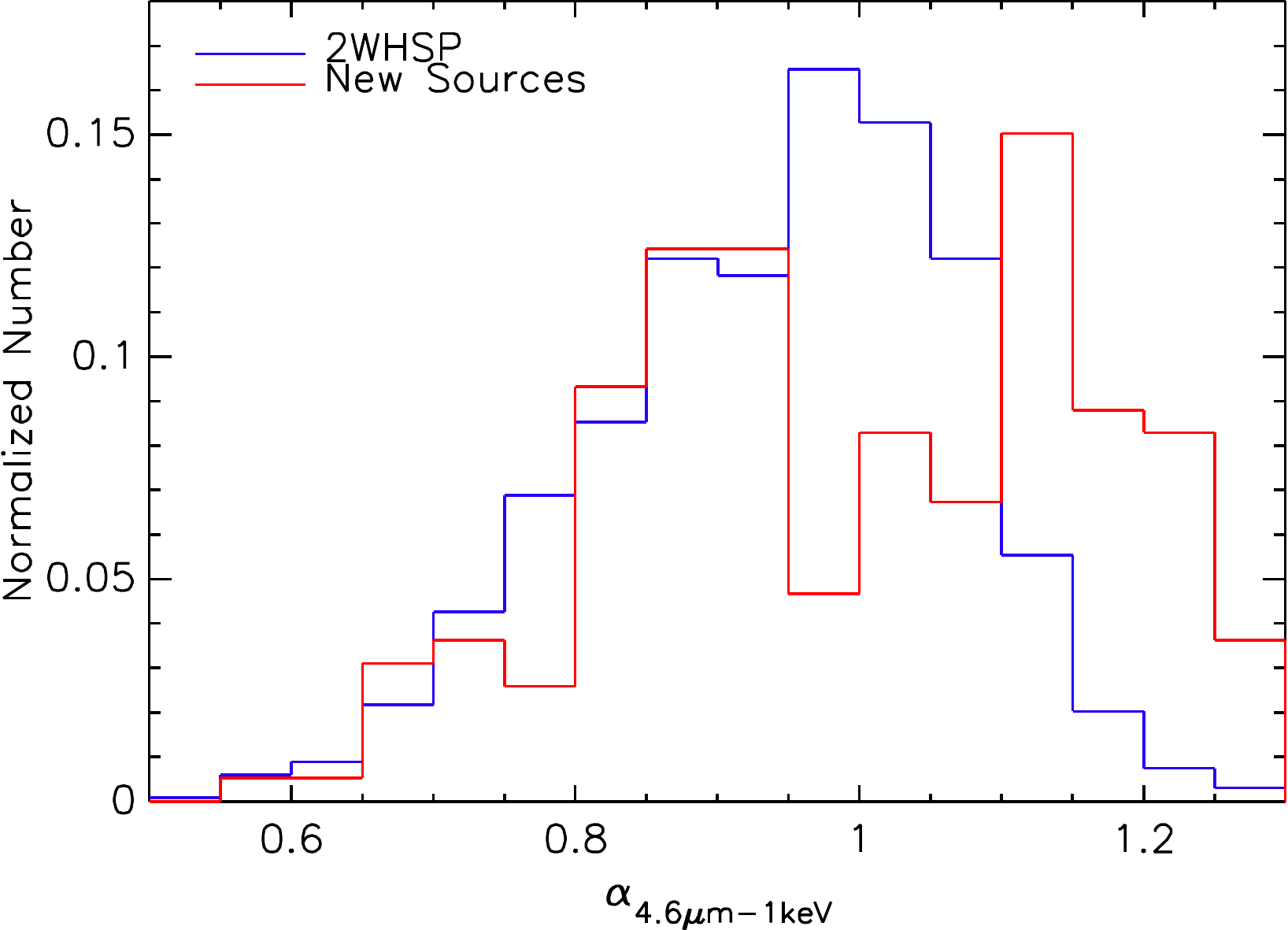}
\includegraphics[width=0.95\linewidth]{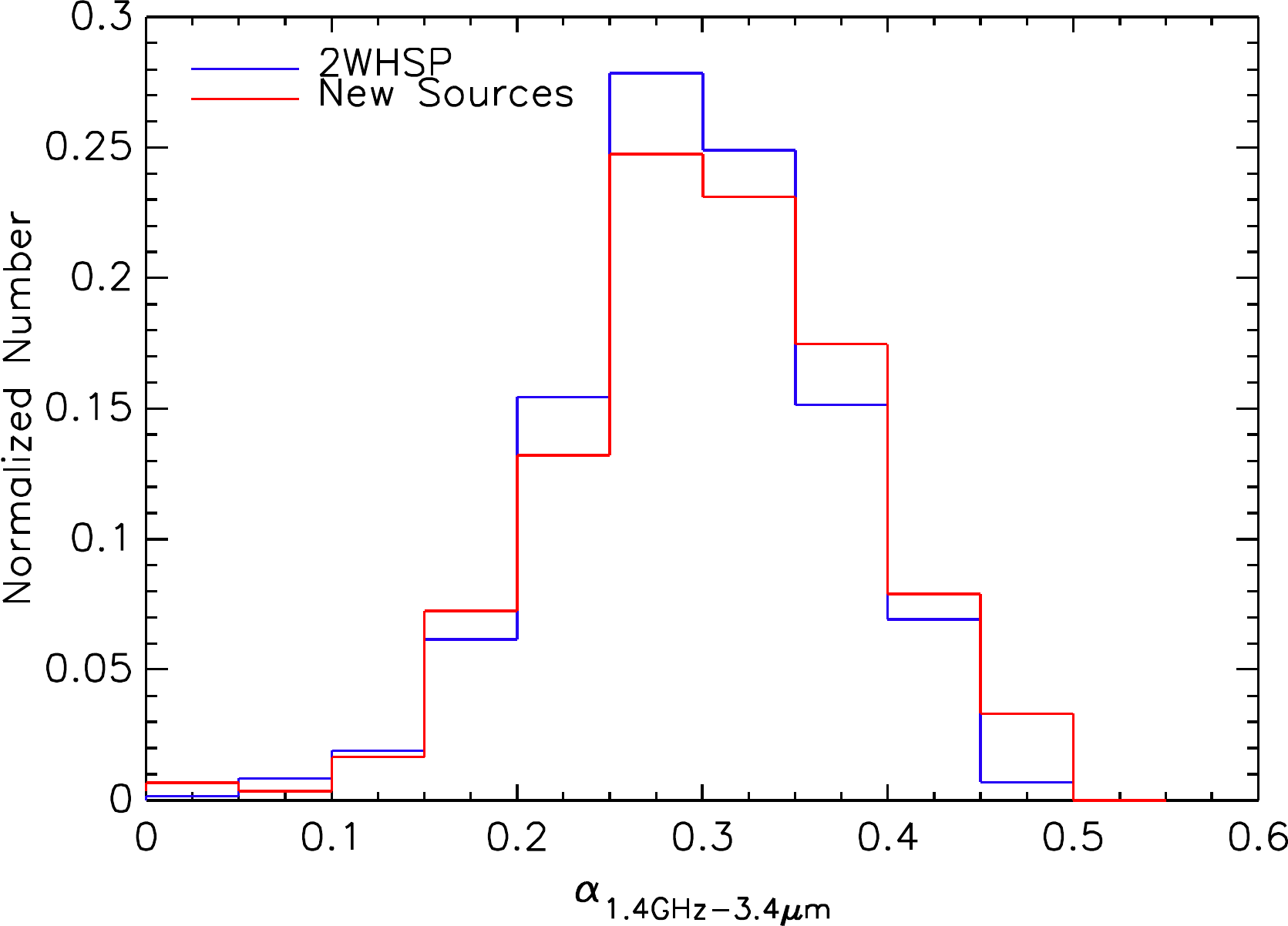}
\caption[The radio-IR slope and the X-ray-IR slope distribution of the new sources and those in the 2WHSP sample.]{Distribution of the new sources and the 2WHSP (X-ray--IR slope, top;  radio--IR
slope, bottom) normalised based on the total number of objects in each sample. }
\label{diffalf}
\end{figure}

A comparison between the new sample and the 2WHSP catalogue is shown in Figs. \ref{diffnumjy} and \ref{diffalf}, normalising on the total number of objects in each sample. 
Figure~\ref{diffnumjy} shows that many of the new sources have \nupeak\ close to the threshold of $10^{15}$ Hz and have higher radio flux compared to the 2WHSP sample. 
Figure~\ref{diffalf} indicates that there are more high $\alpha_{\rm IR-X}$ sources in the new sample, suggesting that relaxing the slope criteria allows the selection of new HSPs with relatively faint X-ray emission. 
The figure also shows that the $\alpha_{\rm r-IR}$ distribution for the subsample of new sources is similar to that of the 2WHSP sample.
This is a strong indication that 2WHSP is close to  complete regarding galaxy dominated HSP sources. 
We have selected almost all the nearby and bright HSPs. 
Around ten sources have $\alpha_{\rm r-IR}$ higher than 0.5, and we examined all of them thoroughly. 
According to their radio morphology and data (1.4 GHz and 5 GHz Planck data), most of them are moderately radio extended or have misaligned jets.
Their IR emission is on average lower than that of  other sources and therefore have relative low IR-to-radio flux ratios.

In conclusion, compared to 2WHSP, the new catalogue includes more sources with \nupeak\ close to the $10^{15}$ Hz limit with brighter radio fluxes. 
By examining all possible blazar counterparts in \gr\ catalogues, we were able to find additional HSP sources with \gr\ emission, which we might have missed if we had only relied on our radio and X-ray selection methods (Sect. \ref{sec:building}). 
We called the new catalogue 3HSP,  removing the `W' as the WISE counterpart is not a requirement in the new catalogue. 

\subsection{Redshift estimations}
\label{redshift}
It is well known that many BL Lacs have no redshift determination because of the lack of any detectable feature in their optical spectra. 
However, for a good fraction of our HSPs, the signature of thermal emission from the host galaxy can be recognised in their broad-band SEDs, especially in the IR band, and this can be used to obtain a photometric redshift (photo-z) estimation.

\citet{Urry2000} showed that the host galaxies of BL Lac objects are giant ellipticals with almost constant absolute magnitude of $\rm{M}_{R}=-23.5$ (see their Fig. 5 for  details). 
By fitting the host galaxy contribution to the SED using a giant elliptical galaxy template \citep{Coleman1980} and assuming that the elliptical galaxy template is approximately a standard candle proxy, with luminosity equal to $\rm{M}_{R}=-23.5$ a photometric redshift can be estimated. 
Therefore, whenever the host galaxy contribution could be distinguished from the non-thermal emission in the SED, we applied this method to estimate the redshifts of all sources with a featureless optical spectrum and those for which no optical spectrum is available.

  
\begin{figure} [h!]
\centering
\includegraphics[width=0.98\linewidth]{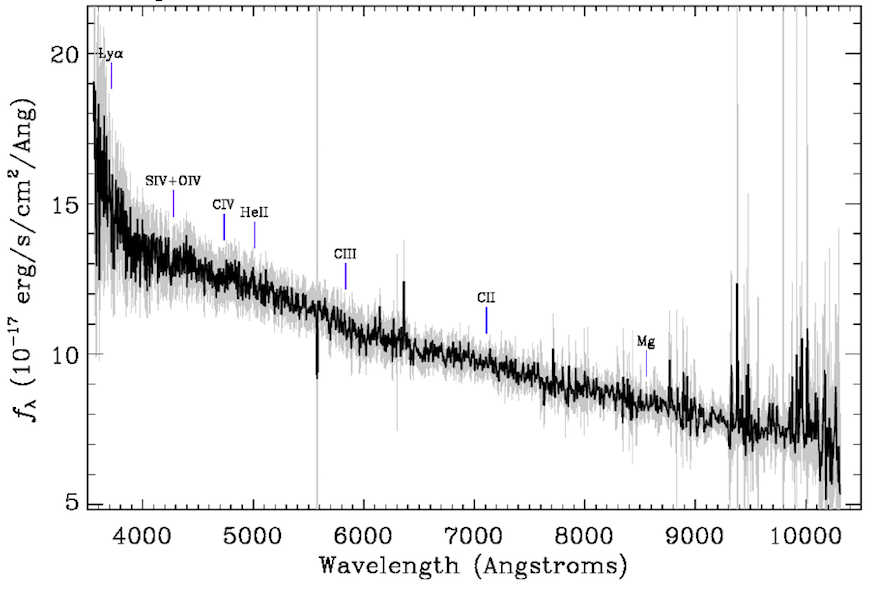}
\includegraphics[width=0.98\linewidth]{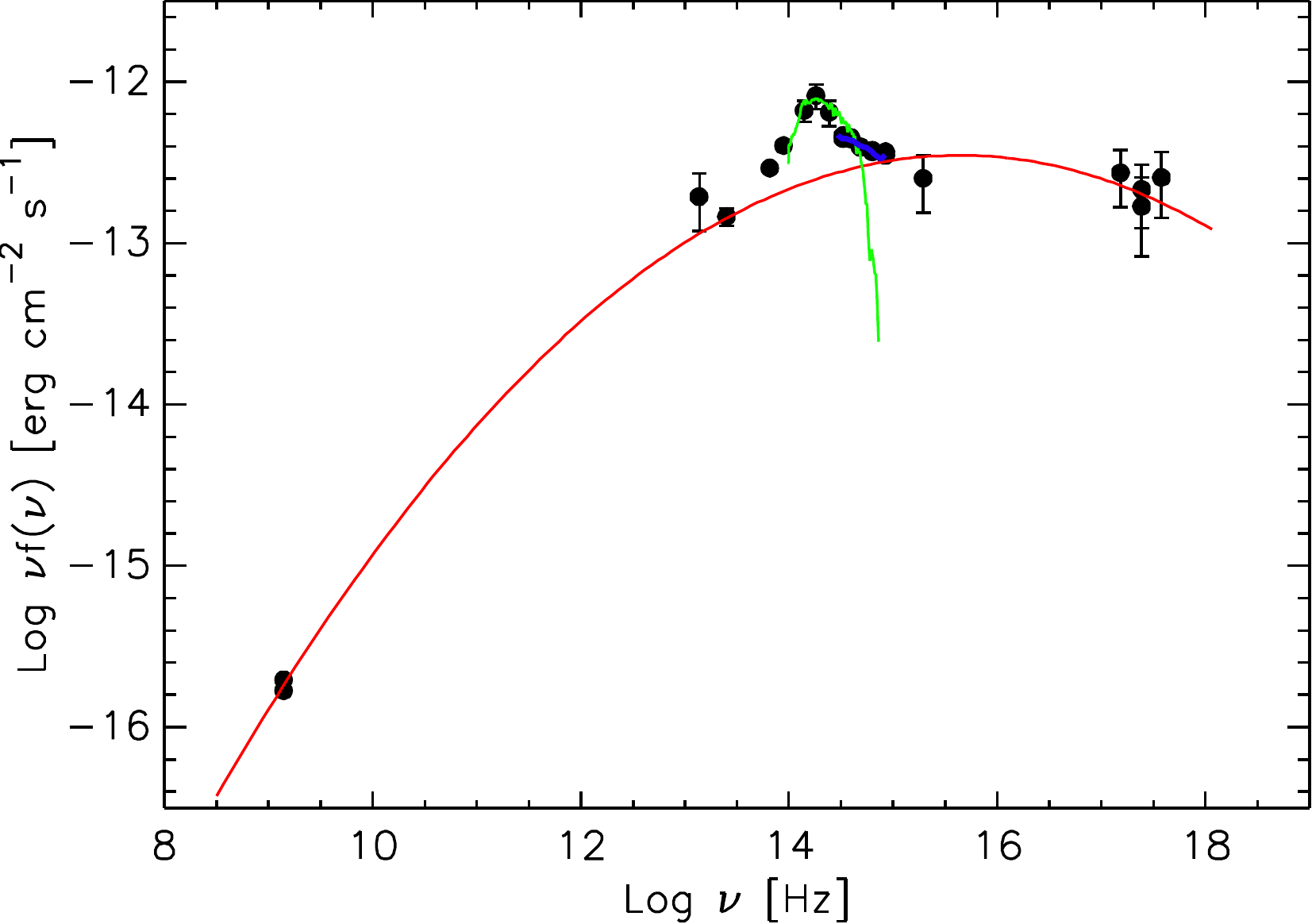}
\caption[illustrated of the photo-z]{Example of photometric redshift (photo-z) estimation. The SDSS Dr14 optical spectrum (top) and SED (bottom) of the source 3HSPJ 154433.1+322148. The red line represents the non-thermal component, the green line  the giant elliptical template fitted to the IR (and partly optical) data at $z=0.32$, and the blue line   the data extract from the SDSS DR14 optical spectrum. }
\label{photozillus}
\end{figure}

An example of the application of this method is shown in Figure~\ref{photozillus} (top left) where    we show the optical spectrum of the source 3HSPJ 154433.1+322148, taken from the Sloan Digital Sky Survey (SDSS) Dr14 \citep[][]{Blanton2017,Abolfathi2017}. 
The spectrum is clearly featureless and no spectral redshift can be estimated. 
In the bottom pane we show the radio to X-ray SED of the same source.
The top panel of figure~\ref{photozillus} shows that the slope of the SDSS spectrum changes above $\approx 7600~{\rm\AA}$. 
According to the SED and the optical spectrum, the SDSS i and z bands are clearly from the host galaxy. 
We then used only the r, g, and u bands in the optical to UV to fit \nupeak.
In addition,  the data extracted directly from the SDSS DR14 optical spectrum  fits well with the optical SED and indicates the same thing.
By fitting the Two Micron All-Sky Survey (2MASS), WISE W1 and W2, and part of the SDSS data using the giant elliptical template available in the SSDC SED builder, we obtained a photo-z estimation of 0.32.

For sources with a featureless spectrum and where non-thermal emission was dominating at all wavelengths, we estimated a redshift lower limit. 
This was done, as in Paper I (1WHSP), by assuming that in the optical band the host galaxy is swamped by the non-thermal emission and leaves no imprint on the optical spectrum when the observed non-thermal flux is at least ten times higher than the host galaxy flux (for details, see Eq. 5 of Paper I). 

\begin{figure} [h!]
\centering
\includegraphics[width=0.98\linewidth]{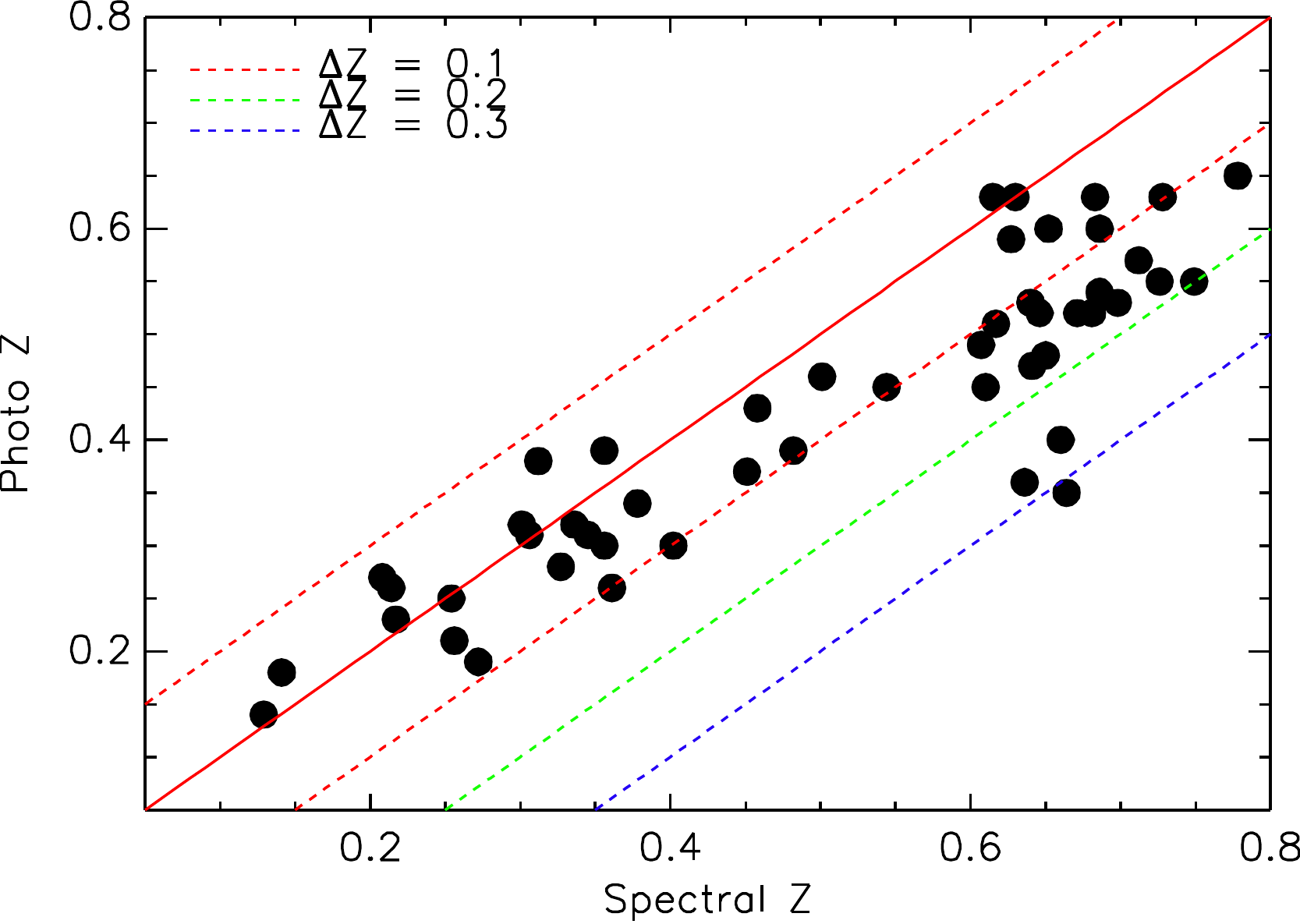}
\caption{Comparison between photo-z and spectral-z. The central solid line has slope of one and  represents the equality between the two redshift estimations, while the red, green, and blue dashed lines indicate a scatter of 0.1, 0.2, 0.3, respectively.}
\label{errorphotoz}
\end{figure}

To estimate the uncertainty of the photometric evaluation of {\it z}, we randomly selected 50 sources with firm {\it z} measurements, and plotted in figure~\ref{errorphotoz} the photometric estimation versus the optical spectra estimation (from SDSS Dr14) of {\it z}.
The figure shows that the differences between the two redshift estimations is at most  0.1 for sources with $z <0.6$, while for $z > 0.6$, the error is at most 0.3.
However, the estimation differences for the majority of these high-z sources is around 0.2. 

There are 80 sources with photo-z higher  than 0.6, and we double checked all of them. 
The error might be large given that the contrast between the host galaxy features and the non-thermal emission is low and not easy to resolve.
The amplitude of the error also depends on the availability of IR--optical data from host galaxy.
For some sources it is more difficult to tell the origin of the IR--optical emission; therefore, they are marked with appropriate source and redshift flags.

Figure~\ref{errorphotoz} illustrates that the photo-z estimation for high-z are underestimated when comparing with spectral-z.
One of the most likely reasons for this bias is due to the difficulty in finding the correct position of the 4000 angstrom break in the optical spectrum. 
For a high-redshift/high-luminosity source the non-thermal flux may be higher than that of the host galaxy at the position of the break. 
It could be that the `real' break is at a lower frequency than where the non-thermal flux merges with the emission from the host galaxy, which is where a slope change can be detected in the optical spectrum. 
That is why we underestimated the photo-z from the SED photometric data.

Recently, a new spectroscopic redshift estimation has become available for 3HSP J062753.2-151956 with spectral-z= 0.31\citep{Paiano2018}, which is very close to our photo-z=0.29.

The breakdown of redshift determinations is as follows:
\begin{itemize}
\item [$\bullet$]  31.8\% of 3HSPs with firm redshift (flag 1);
\item [$\bullet$]  5.3\% of 3HSPs with uncertain redshift (flag 2);
\item [$\bullet$]  4.7\% of 3HSPs with lower limit redshift (flag 3);
\item [$\bullet$]  7.2\% of 3HSPs with photometric redshift and featureless optical spectrum (flag 4);
\item [$\bullet$]  39.0\% of 3HSPs with photometric redshift and without optical spectrum (flag 5);
\item [$\bullet$]  11.9\% of 3HSPs without any redshift estimation or measurements.
\end{itemize}

\begin{figure} [h!]
\centering
\includegraphics[width=0.98\linewidth]{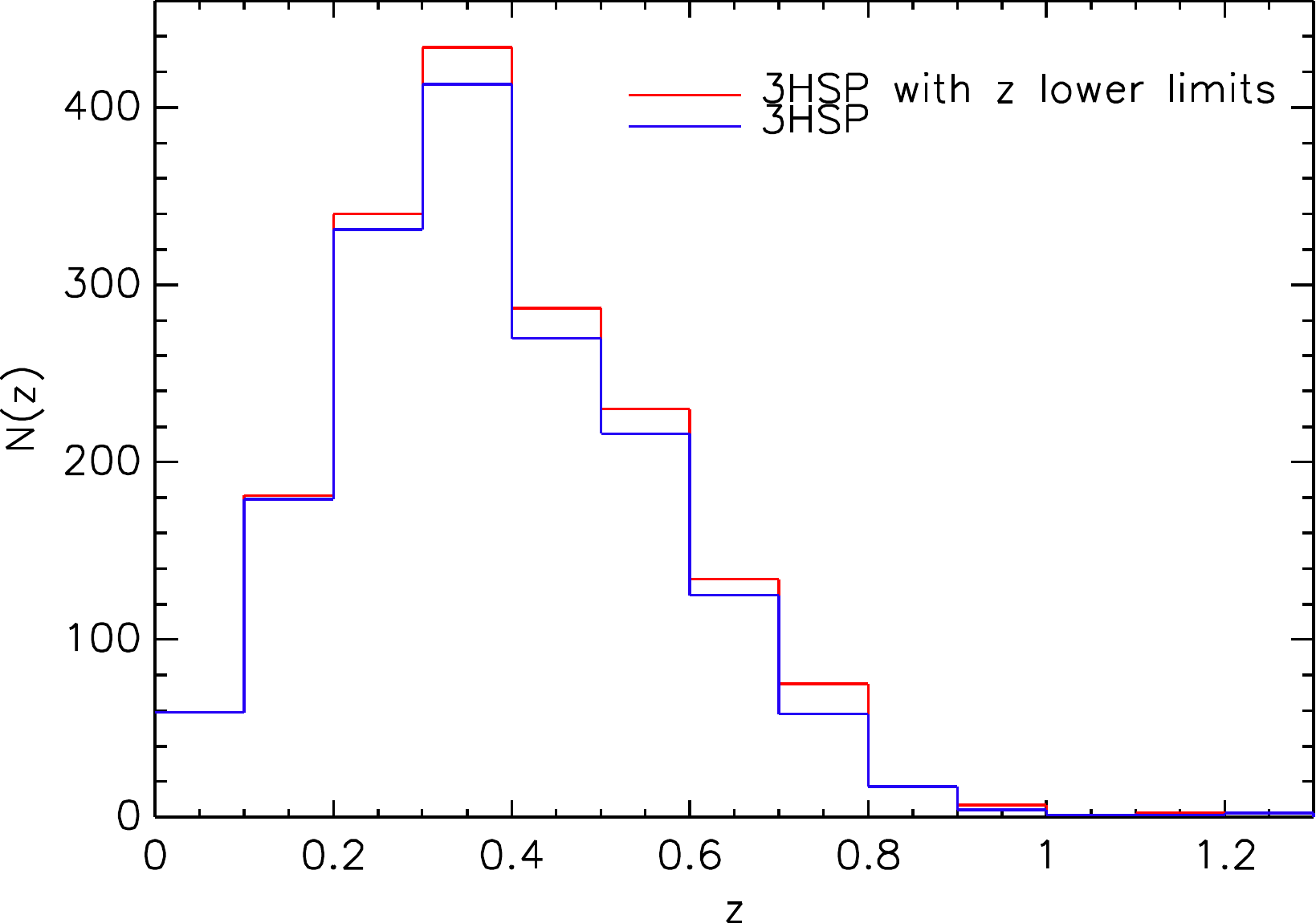}
\caption{Redshift distribution of the 3HSP sample. The red line represent all the 3HSP sources with a redshift estimation, while the blue line represents sources with an estimation, but not lower limits.}
\label{zdist}
\end{figure}

In Figure~\ref{zdist}, we show the redshift distribution for the entire 3HSP catalogue. 
The figure suggests that the redshift of 3HSP sources is centred around ${z=0.35}$, which is similar to previous results (e.g.  1/2WHSP, 5BZCat, Sedentary Survey).

The redshift of the sources without redshift might be significantly higher than the average value. 
According to the Monte Carlo simulations of \citet{Giommi2012}, the average predicted redshift of BL Lacs with featureless spectra is around 1.2. 
Figure 9 of the 2WHSP paper also suggests that HSPs with featureless spectra might be much more distant than we think.  

\subsection{XRT data analysis with the {\it Swift} Deepsky pipeline}
\label{xrtanalysis}
During the selection of 3HSP objects, we proposed {\it Swift} observations for 210 sources without a good X-ray spectrum (50 from 1WHSP, 80 from 2WHSP, and 80 from 3HSP), 190  of which were kept in the 3HSP final version. 
As of March 2018, 151 of these sources have been observed by {\it Swift}. 
We  analysed all the XRT data using the {\it Swift} DeepSky software\footnote{https://github.com/chbrandt/swift\_deepsky}\citep{giommi2019}, a pipeline tool based on HEASoft6.25 and the XIMAGE package that we  assembled and that can run on a Docker container\footnote{https://hub.docker.com/r/chbrandt/swift\_deepsky/} \citep{Morris2017}. 
This software detects sources and estimates fluxes (or upper limits) in four energy bands. 
For the sources detected with at least 100 photons we used XSPEC to estimate a best fit spectrum using power law and log parabola models.  

Our analysis resulted in 147 new X-ray detections in association with 3HSP sources. 
All new X-ray data, together with optical and UV measurements from {\it Swift} Ultraviolet/Optical Telescope (UVOT), were used to estimate or update \nupeak (analysed using the SSDC on-line interactive analysis tool). 
The high {\it Swift} detection rate clearly shows that the use of multi-frequency information for selecting X-ray targets is very effective for the detection of blazars in that band; the only four non-detections ($<1$\%) are objects observed with very short exposures. 

\section{3HSP: the largest and most complete catalogue of HSP blazars to date}
\label{catalog}

\begin{figure} [h!]
\centering
\includegraphics[width=0.95\linewidth]{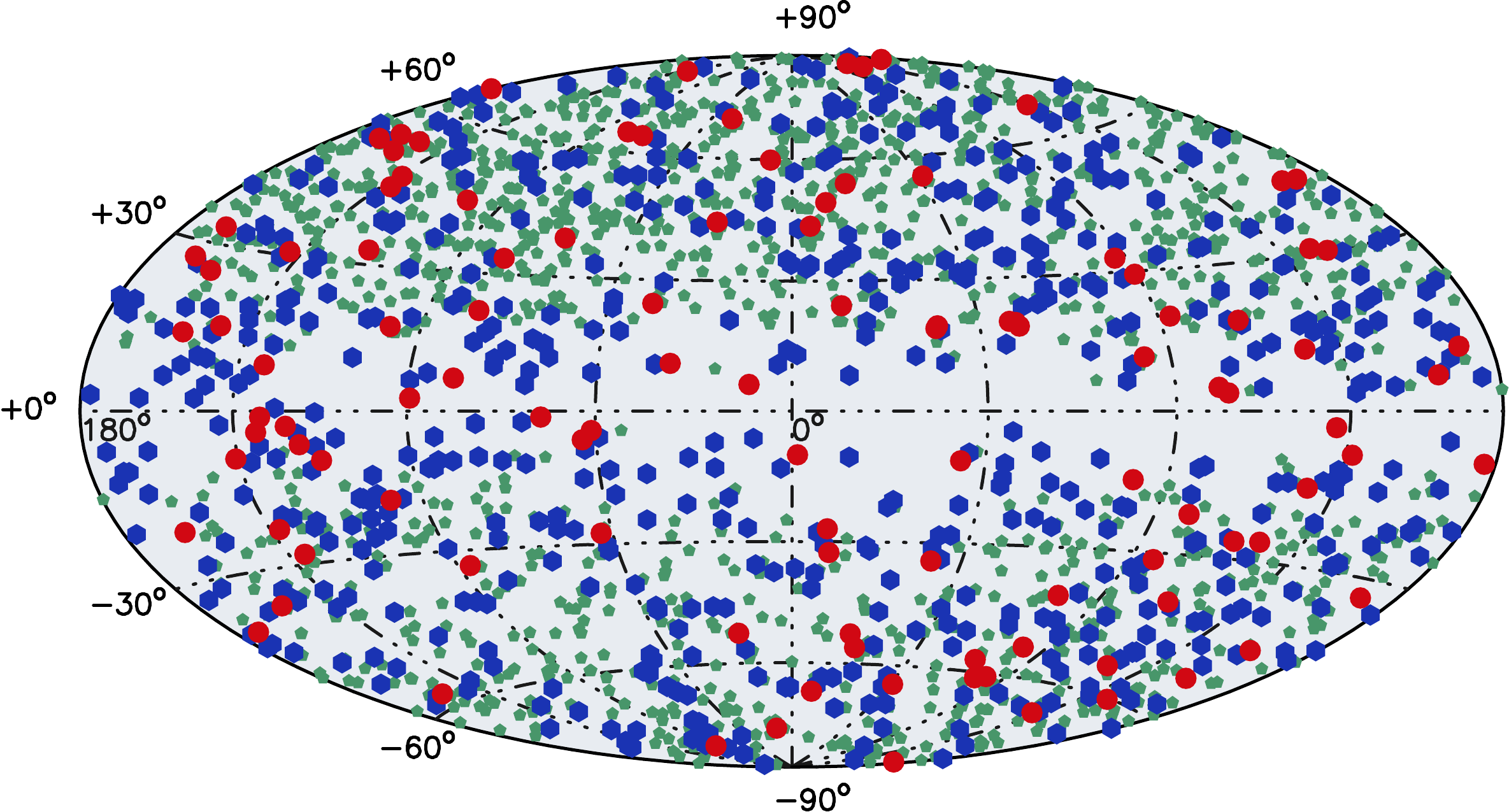}
\caption[Aitoff projection of 3HSP.]{Aitoff projection of 3HSP sources in Galactic coordinates. Red, blue, and green points represents ${\rm FOM} > 1.0$,  $0.2 < {\rm FOM} < 1$, and ${\rm FOM} < 0.2$, respectively (see text for details).} 
\label{3hspview}
\end{figure}

In total, the new version of the catalogue includes 2013 sources, 88 of which are close to the Galactic plane ($|b|<10^{\circ}$). 
This catalogue has been named 3HSP, which means the third catalogue for HSP blazars. 
The W in the acronym has been removed in this edition of the catalogue as sources are no longer required  to be detected in the WISE all sky survey.
Of the 2013 3HSP sources, 1618 are also in 2WHSP, 657 are in 5BZCat, and 1007 have a counterpart in one of the {\it Fermi}-LAT catalogues. 
Figure~\ref{3hspview} illustrates the Aitoff projection map of the 3HSP catalogue in Galactic coordinates, and most of the 3HSP sources are located out of the Galactic plane.  

The content of the 3HSP catalogue presented in this paper differs from 2WHSP as detailed below:
\begin{itemize}
    \item  it includes 395 new HSP blazars/candidates; 
    \item  73 sources that were in the 2WHSP catalogue have been removed, due to reclassification as intermediate synchrotron peaked (ISP),  low-energy synchrotron peaked (LSP) objects,\footnote{In ISP blazars $\nu_{\rm peak}$ is between $10^{14}$ and $10^{15}$~Hz, and in LSP $\nu_{\rm peak} < 10^{14}$~Hz} or spurious associations based on new optical spectra and XRT data;
    \item we added sources mainly based on the radio/X-ray flux ratio and \gr\ catalogues;
    \item photo-z values were estimated for 930 sources; 
    \item appropriate flags were assigned to uncertain sources.
\end{itemize}

We summarise the steps followed to build the 3HSP catalogue here: 
\begin{enumerate}
    \item Cutting the NVSS-RASS matched sources with flux ratio $\geq 9\times10^{-11}~\rm{erg}~\rm{s}^{-1}~\rm{cm}^{-2}~\rm{Jy}$ and excluding those already in 1/2WHSP selection process, leading to 3011 pre-selected sources; 
    \item Cross-matching the pre-selected sources with 5BZCat, XMMOM, and Fermi 3FHL, reducing the pre-selected sample to 254 sources; 
    \item Examining  the pre-selected candidates one by one, adding new 58 HSPs and HSP-candidates to the catalogue;
    \item  Checking additional blazars from the Fermi 3FHL catalogue (168 sources) and Fermi 4FGL catalogue (121 sources); 
    \item Cleaning non-confirmed sources, especially those with no Fermi counterpart, removing 73 sources previously listed in 2WHSP; 
    \item Adding 48 new sources identified using the VOU-Blazars tool. 
\end{enumerate}

Table~\ref{whole3hsp} lists the \nupeak, redshift, \gr\ counterpart, 2WHSP counterpart, BZCat counterpart, and a Figure of Merit (FOM) of 
a representative subsample. 
The full catalogue is available in electronic form\footnote{\url{www.ssdc.asi.it/3hsp/}}.
The FOM was defined in \citet{Arsioli2015} as the ratio of the flux at the synchrotron peak (\nufnupeak) of a given source to the peak flux of the faintest 1WHSP blazar that had been detected in the TeV Band. 
Here we re-define the FOM in units of $2.5\times10^{-12}~{\rm erg}~{\rm s}^{-1}~{\rm cm}^{-2}$ (Log \nufnupeak~$=-11.6$), which is the peak flux of the faintest 3HSP source in the current version of TeVCat. 
This parameter was introduced to provide a simple quantitative measure of potential detectability of HSPs by TeV instruments (see Paper I for more details).


Some of the 3HSPs sources have sparse non-thermal data, poor quality data, or  peculiar flux ratios. These sources will need to be carefully followed up in future versions of the HSP catalogue. 
These objects were assigned a flag value that reflects the reason for the uncertainty. 
For sources with little non-thermal data or with fairly large X-ray positional errors, we gave a flag value of 1. 
Flag 2 is for cases with one or two doubtful data points (due to large positional uncertainty or problematic photometry) but that still provide a reasonable HSP-like SED.
Sources with a low ratio between synchrotron peak flux and radio flux (possibly due to jet misalignment) were given source flag 3.
Source flag 4 is for cases where the observed IR or optical emission cannot be safely attributed to the host galaxy or to synchrotron emission. 

One of the main purposes of the 3HSP catalogue is to provide possible seed sources for future HE and VHE observations, so we would like to select as many candidates as possible. 
There might be a small fraction of sources that still need further data to be confirmed as HSPs. 
Among those with an uncertain flag, sources with flag 1 are the most ambiguous, but we have no reason to remove them from the current version of the catalogue (some of them even show a blazar-like optical spectra), although we highlight the need of a follow-up to confirm their classifications. 
  
There are 374 sources (18.6\%) assigned with flag 1 in the 3HSP 
catalogue, which implies an efficiency of at least 81.4\%. 
We note that less than 5\% of the sources in 2WHSP turned out to be spurious HSPs. Therefore, given that we now  have access to more data and better tools, we expect a lower rate of spurious classifications for the 3HSP catalogue. 
A more detailed estimation of the number of spurious sources is given in section~\ref{spurious}.
In addition, we have a good record of selecting HSP-candidates for further X-ray observations, especially with {\it Swift}. 
We have already had three successful {\it Swift} observation campaigns carried out based on 1/2WHSP sources, resulting in ~200 dedicated observations to HSPs and HSP-candidates.

Flags are also associated with \nupeak\ and redshift values with the following meaning: 
\begin{enumerate}
    \item firm estimation;
    \item uncertain value;
    \item lower limit;
    \item photometric redshift of an object with featureless optical spectrum;
    \item photometric redshift of an objects for which no optical spectrum is available. 
\end{enumerate}
We note that flags 4 and 5 only apply to redshift.
Moreover, the uncertain value for the \nupeak\ estimate means that we are not sure exactly what   the synchrotron peak frequency is, due to insufficient non-thermal data, but we could still tell that the frequency is higher than $10^{15}$ Hz.
Sources with an uncertain flag for synchrotron peak frequency are not necessarily candidates, as some of them have blazar like optical spectrum or have  already been included in 5BZCat. 
The source flag and the synchrotron peak frequency flag are marked independently. 
 
As shown in Figure~\ref{photozillus}, sources with a redshift flag equal to 4 have featureless optical spectra, but the emission from the host galaxy is not completely overwhelmed by the non-thermal radiation, thus a photo-z can still be estimated from IR data or in part of the optical band. 
Flag 3 (lower limits) sources also have   featureless spectra; however, their SEDs are totally dominated by synchrotron radiation. 
There are still some sources with SEDs that are non-thermal dominated and for which no optical spectrum is available. The redshift in this case remains blank; this applies only to $11.94\%$ of the sources. 



\section{Completeness and statistical properties of the 3HSP catalogue}
The demographic properties and the cosmological evolution of blazars have been extensively debated by the community, and for a long time the existence of the so-called blazar sequence has been a controversial topic. 
With the largest ever HSP blazar catalogue, the overall properties of the 3HSP sample can be discussed thoroughly, and here we present arguments in tension with the blazar sequence scenario. 
We start by checking the completeness of the 3HSP catalogue in the radio and the X-ray energy bands. 
\begin{figure} [h!]
\centering
\includegraphics[width=0.98\linewidth]{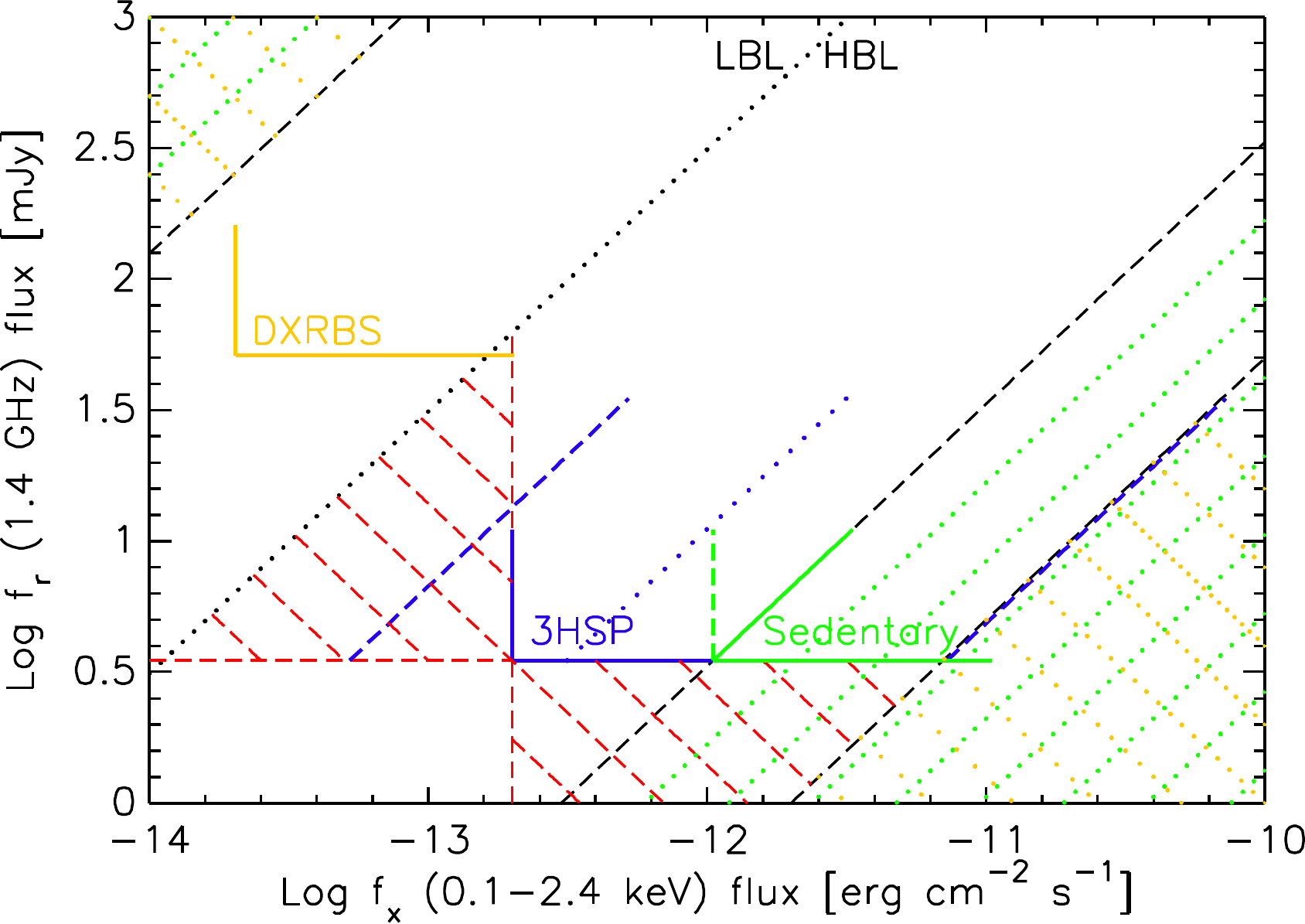}
\caption[The sampling of the radio flux density - X-ray flux plane]{Sampling of the radio flux density--X-ray flux plane with the 3HSP, Sedentary, and DXRBS samples. The blue dashed and dotted lines are the slope limits for the 3HSP sample. The black dotted line is the flux ratio that separates LBL and HBL. The green and yellow dotted lines denote the forbidden region for BL Lacs and  FSRQs.  The red dashed lines indicate the incomplete regions of 2WHSP. }
\label{rxfig}
\end{figure}

Figure~\ref{rxfig} shows the radio--X-ray flux plane of the 3HSP, comparing it with the Sedentary survey \citep{Giommi1999,Giommi2005,Piranomonte2007} and the Deep X-ray Radio Blazar Survey \citep[DXRBS][]{Padovani2007a}.
The corresponding approximate synchrotron peak frequencies were estimated using extrapolation, and the relationship between \nupeak\ and the X-ray--to--radio flux ratio (${\rm f_x/f_r}$) with the equation, \nupeak~$=~({\rm f_x/f_r}~+~16.068\pm0.306~)~/~0.377\pm0.019$.
According to the error of the two fit parameters, we estimate that the uncertainty on this estimation is around one order of magnitude for \nupeak.
We note that this relationship was derived from the 3HSP subsample, so we only convert the radio--to--X-ray flux ratio to synchrotron peak frequency for HSPs (or equivalently \nupeak$>10^{15}$~Hz).

The black dotted line indicates the X-ray--to--radio flux ratio that separates low-energy peaked BL Lacs (LBL) and HBL, which is ${\rm f_x/f_r}=3.2\times10^{-12}~{\rm erg}~{\rm cm}^{-2}~{\rm s}^{-1}~{\rm Jy}^{-1}$ in \citet{Padovani2003} and \citet{Padovani2007}. 
Three black dashed lines represent the flux ratios, from left to right, ${\rm f_x/f_r}=8\times10^{-14}, 3\times10^{-10}$, and $2\times10^{-9}~{\rm erg}~{\rm cm}^{-2}~{\rm s}^{-1}~{\rm Jy}^{-1}$, respectively. 
The first value represents the minimum X-ray--to--radio flux ratio for  BL Lac objects and 
for flat spectrum radio
quasars (FSRQs) \citep{Padovani2007} based on the data from \citet{Padovani1997} and \citet{Siebert1998}.
The second value is the maximum flux ratio for FSRQs, while the third  is the maximum flux ratio for BL Lacs (\nupeak~$\approx10^{20}$~Hz: \citealt{Padovani2003,Padovani2007}).

Among the radio and X-ray catalogues used for the 3HSP selection, NVSS and RASS have the largest sky coverage, and we used the NVSS and RASS catalogues to estimate the radio and X-ray limits for Fig.~\ref{rxfig}. 
Given that the minimum radio flux cut applied for the 3HSP--NVSS subsample, the radio limit for this 3HSP subsample in this figure is set to 3.5~mJy, while the X-ray limit for the  3HSP-RASS subsample  is set to the minimum RASS flux value in the subsample, $\approx 2 \times 10^{-13}~{\rm erg}~{\rm cm}^{-2}~{\rm s}^{-1}$. 
However, it should be noted that in some cases radio and X-ray flux limits can be lower  as we used several other radio and X-ray catalogues to build the 3HSP sample. 
In practice, the exact radio and flux limits of 3HSP are lower than the limits set here. 

According to Fig.~\ref{rxfig} the 3HSP is not complete, neither in  radio nor in X-rays.
Sources with radio flux brighter than 3.5~mJy but X-ray flux fainter than $2 \times  10^{-13}~{\rm erg}~{\rm cm}^{-2}~{\rm s}^{-1}$ (upper red dashed region) are not selected as they are not detected by current large-area X-ray surveys. 
On the other hand, some bright X-ray but faint radio sources (lower red dashed   region)  are missed, since these sources are not included in today's large-area radio catalogues. 
If we were to increase the radio flux limit to $\approx22$~mJy, we would define a complete, radio flux-limited HBL sample. 
Similarly, when setting the X-ray flux limit to a higher value $\approx7\times~10^{-12}~{\rm erg}~{\rm cm}^{-2}~{\rm s}^{-1}$, the sample would become complete in the X-ray band. 

The blue dotted line in the figure represents the radio to X-ray ratio ${\rm f_{\rm x}/f_{\rm r}}=9\times10^{-11}~{\rm erg}~{\rm cm}^{-2}~{\rm s}^{-1}~{\rm Jy}$ , which is the flux criteria applied when adding new sources that are not in the 2WHSP catalogue.
This value is just slightly lower than the average value we obtained from the old 2WHSP subsample with \nupeak\ close to $10^{15}$~Hz. 
The left blue dashed line (${\rm f_{\rm x}/f_{\rm r}}=1\times10^{-11}$) corresponds to the slope criteria applied when building the 1/2WHSP, and yields a relatively low  \nupeak\ value ($10^{14.2}$~Hz), while the blue dotted line marks the sources with \nupeak\ around $10^{15}$~Hz.
Some HSPs have a radio--to--X-ray flux ratio lower than ${\rm f_{\rm x}/f_{\rm r}}=9\times10^{-11}$, and to select them during the assembling of previous HSP catalogues (1/2WHSP) we applied a selection criterion corresponding to lower \nupeak\ sources. 

The `lost sources' in the upper red dashed region of Fig. 9 are mainly lower \nupeak\ blazars, while those in the lower  region are essentially higher \nupeak\ blazars.
The consequences of the slope criterion applied\footnote{$0.05<\alpha_{1.4{\rm GHz}-3.4\mu{\rm m}}<0.45,0.4<\alpha_{4.6\mu{\rm m}-1{\rm keV}}<1.1$} \citep{Arsioli2015} are the blue dashed lines displayed, and the estimated lower and upper limits of ${\rm f_x/f_r}$ with this criterion are $1.490 \times 10^{-11}~{\rm erg}~{\rm cm}^{-2}~{\rm s}^{-1}~{\rm Jy}$ (\nupeak\ $\approx 10^{14.2}$~Hz) and $2.063 \times 10^{-9}~{\rm erg}~{\rm cm}^{-2}~{\rm s}^{-1}~{\rm Jy}$ (\nupeak\ $\approx~10^{20}$~Hz), respectively. 
The two slope  limits are estimated with the assumption that the average X-ray spectral slope is 0.9 and the IR slope is 0.3. 

Figure~\ref{rxfig} shows that only by setting a double flux limit (both at radio and X-rays) do we manage to select a 3HSP subsample that avoids the red areas, and that is therefore  complete. 
Thus, we built a statistically complete sample by applying both radio and X-ray cuts. 
Among the radio and X-ray catalogues used for the 3HSP selection, NVSS and RASS have the largest coverage, thus all the statistical tests are based on the subsample of 3HSP sources with both NVSS and RASS counterparts. 
We built a subsample of sources detected both by NVSS and RASS, which we name 3HSP-NVSS-RASS, to be used for statistical studies. It has a cross-matching radii of 0.3 arcmin and 0.8 arcmin, respectively, and includes a total of 1247 sources.

The sensitivity of a survey, as for the case of the ROSAT survey, may not be the same everywhere on the sky. 
In this particular case only a small fraction of the sky is actually observed with the highest sensitivity. 
Here and in the following sections we take this into account by weighing the area of sky available based on the X-ray flux of each source \citep{Voges1999,Voges2000}. 

\begin{figure} [h!]
\centering
\includegraphics[width=0.98\linewidth]{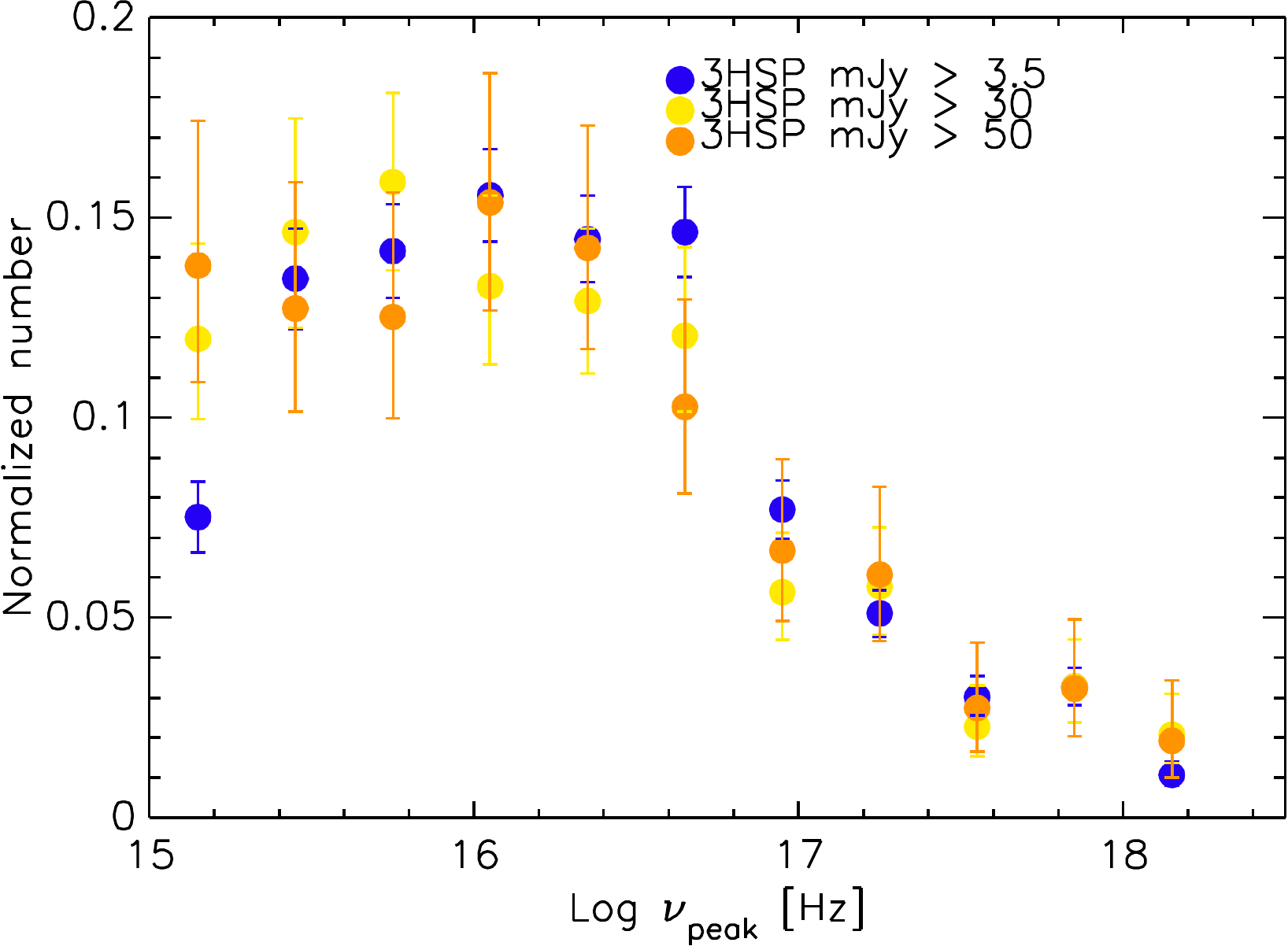}
\includegraphics[width=0.98\linewidth]{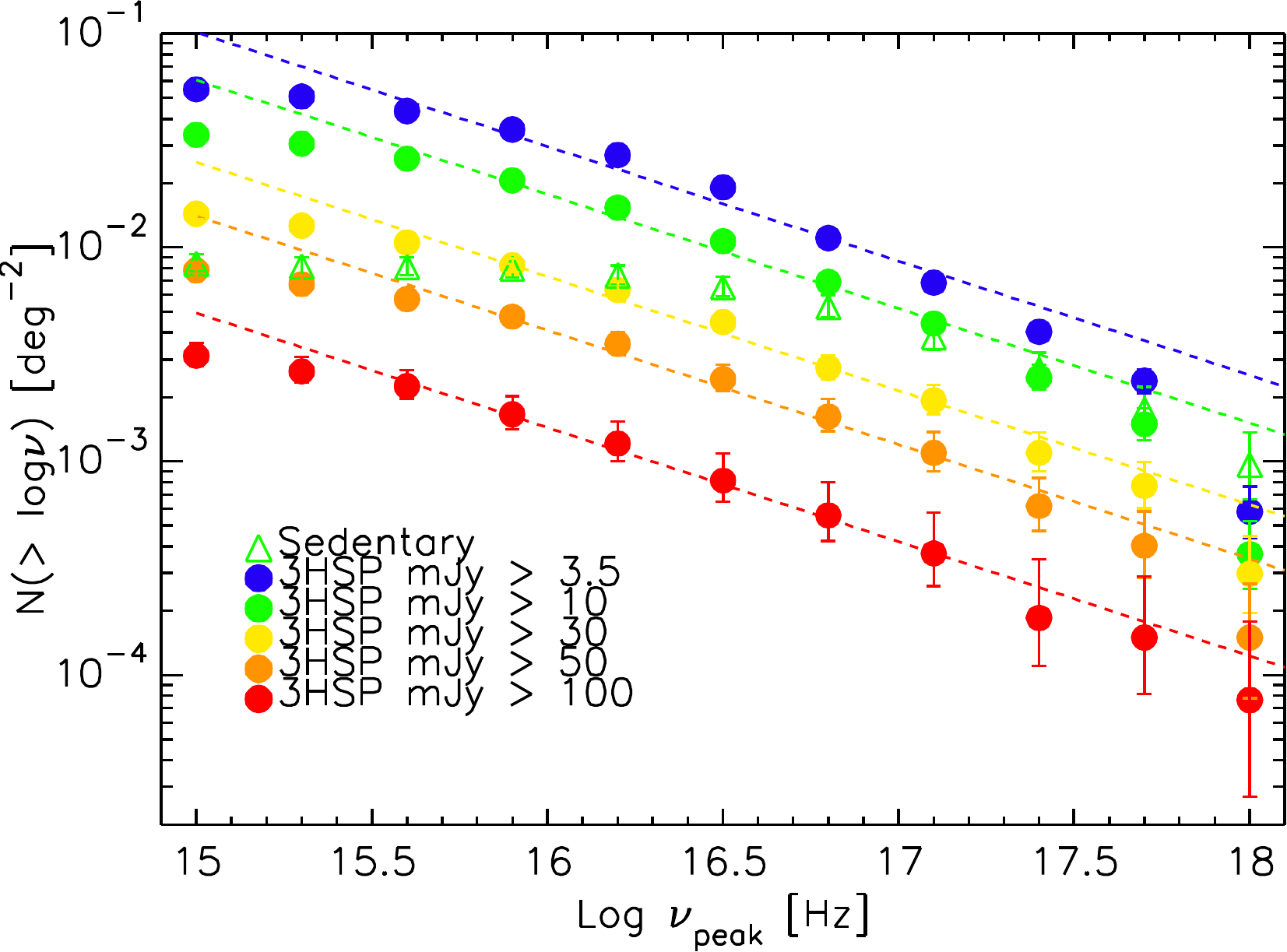}
\caption{Differential (top) and cumulative (bottom) synchrotron \nupeak\ distribution for the 3HSP-NVSS-RASS sample with different radio flux density cuts. Different colours indicate different radio flux cuts. The dashed lines represent the best fixed-slope linear fitting for each radio-cut subsample.}
\label{nudistcalb}
\end{figure}

Figure~\ref{nudistcalb} illustrates the X-ray-weighted synchrotron \nupeak\ distribution for the 3HSP-NVSS-RASS subsample with different radio flux density cuts. 
This figure suggests that the 3HSP-NVSS-RASS sample is complete for \nupeak\ $\gtrsim 10^{16}$~Hz;   for all radio flux density cuts the distribution flattens at the low end \nupeak\ $\lesssim  10^{16}$~Hz, and for the 3.5 mJy cut the data suggests incompleteness for  \nupeak\ $\lesssim  10^{16}$~Hz. 

The cumulative \nupeak\ distribution, compared with the radio complete Sedentary catalogue (green triangles), are shown in Figure~\ref{nudistcalb}. 
The dashed lines, representing the standard slope for a non-evolving complete sample, show a likely incompleteness at low \nupeak\ and low radio flux density values. 
However, even for a 50 mJy cut the distribution tends to flatten at the low end; there are still several bright radio sources with \nupeak\ close to the selection threshold that might be missed due to their faint X-ray flux. 

\begin{figure} [h!]
\centering
\includegraphics[width=0.98\linewidth]{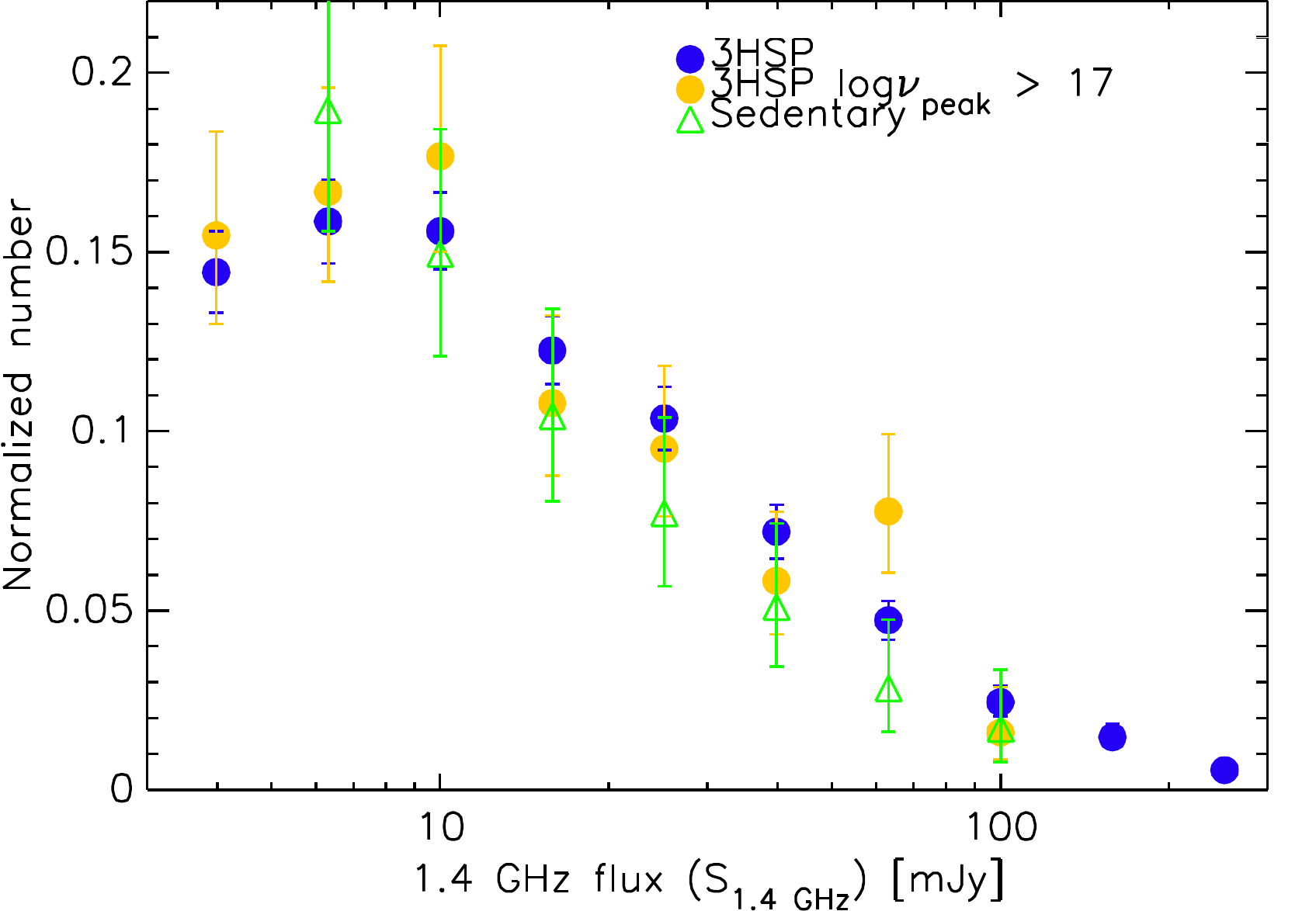}
\caption{Radio flux density distribution with different \nupeak\ cuts.}
\label{mjydist}
\end{figure}

Figure~\ref{mjydist} shows the distribution of radio flux density with different \nupeak\ cuts compared to that of the Sedentary survey \citep{Giommi1999}. 
It looks like the 3HSP-NVSS-RASS subsample is very close to being complete for ${\rm F_{1.4~GHz}} \gtrsim 10$~mJy; however, compared with the distribution from Sedentary, the slope is flatter.
There are still some sources with ${\rm F_{1.4~GHz}}$ between 10 to $\approx$~25~mJy lost from the X-ray selected sample, implying that the 3HSP subsamples are complete only for radio flux cuts brighter than 25 mJy.
The figure also shows a flattening with higher \nupeak\ values for faint radio sources; therefore, those relatively high \nupeak\ sources might be missing due to evolution. 
Since there was a radio cut already applied for the 3HSP-NVSS-RASS subsample and there was no IR constraint when selecting the 3HSP sources, we do not expect to lose high \nupeak\ sources. 

\begin{figure} [h!]
\begin{center}
\includegraphics[width=0.98\linewidth]{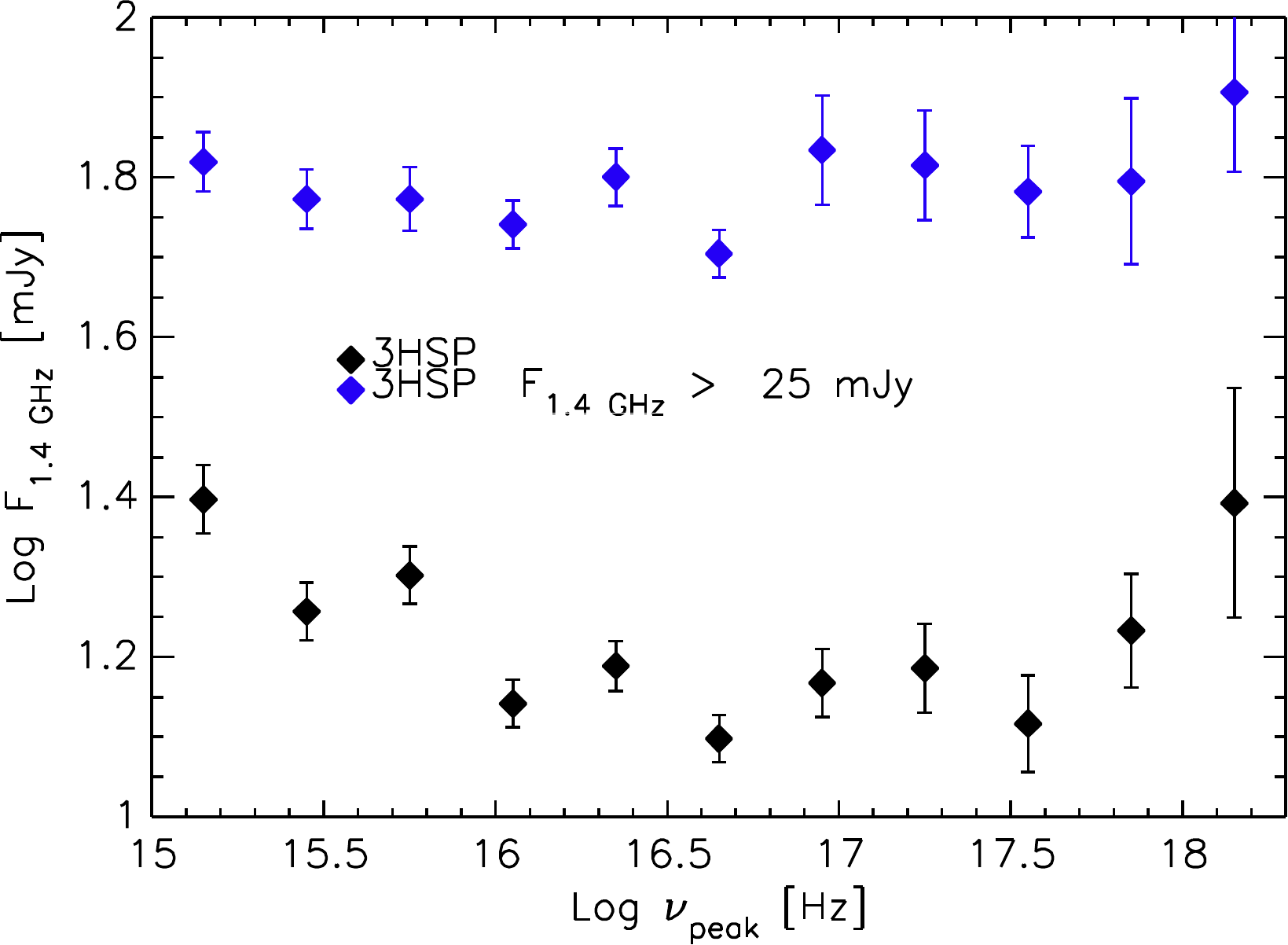}
\end{center}
\caption{Radio flux density vs. \nupeak\ for the entire 3HSP sample (black), and for a complete 3HSP subsample with radio flux cut at 25 mJy (bue).}
\label{mjynu}
\end{figure}

The average radio flux density with respect to the synchrotron \nupeak\ is shown in Figure~\ref{mjynu}, suggesting that the radio flux density does not depend on the synchrotron peak. 
Without the radio cut, the average radio flux density is slightly higher on both ends, probably resulting from the incompleteness and evolution of the HSP blazars. 
For high \nupeak\ sources, we are probably missing faint radio flux sources because they might evolve negatively, 
while for sources with \nupeak\ close to the selection threshold, this is likely due to incompleteness.
For the subsample with a radio flux limit of 25 mJy (blue points in  Figure~\ref{mjynu}) it is clear that the radio flux density is independent of \nupeak, indicating that the fraction of HSP among blazars is independent of radio flux densities. 

\begin{figure} [h!]
\centering
\includegraphics[width=0.98\linewidth]{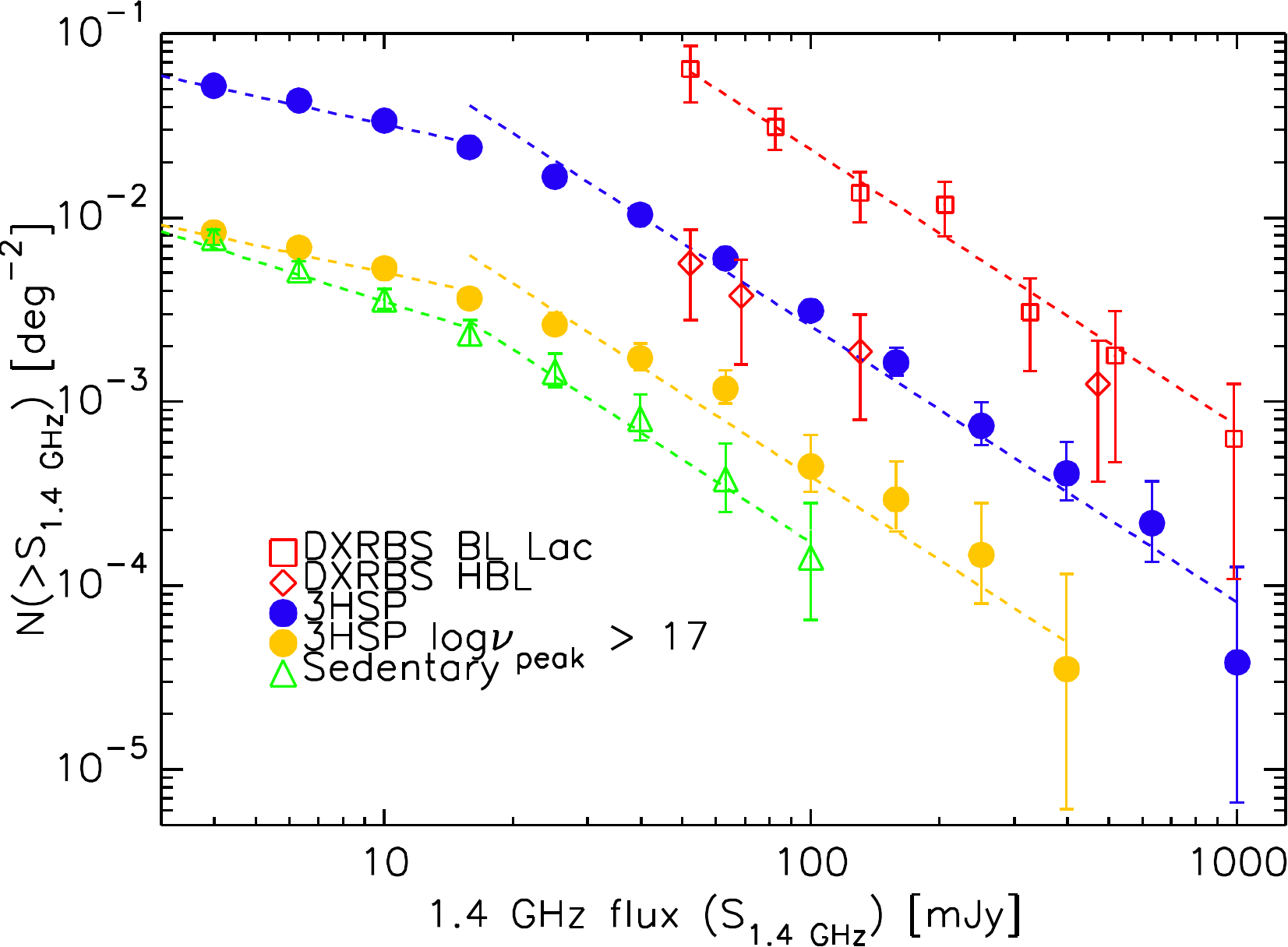}
\caption{Radio logN-logS of different samples of blazars. Blue circles represent the 3HSP sample, green triangles the Sedentary sample, red squares the subsample of BL Lacs from DXRBS, red trapezoids   the subsample of HSPs from DXRBS, and orange circles   the 3HSP subsample with log($\nu_{peak}$)~$>$~17.}
\label{lognsrr}
\end{figure}

A good way to check for the completeness of a catalogue is to evaluate the logN-logS.
The integral radio $\log$N-$\log$S for the 3HSP-NVSS-RASS subsample with different \nupeak\ cuts is shown in Fig.~\ref{lognsrr}. 
Number counts for Sedentary HBL, DXRBS BL Lacs, and DXRBS HBL only are shown as well for comparison. 
The $\log$N-$\log$S for DXRBS are at 5 GHz; however, given that BL Lacs typically have radio spectra with $\alpha_r \sim 0$ \citep{Giommi2012b,Caccianiga2001}, no conversion between the 5 GHz counts and 1.4 GHz is necessary. 
The dashed lines in bright bins correspond to a fixed slope of -1.5, the expected value for a complete sample of a non-evolving population in a Euclidean Universe. 
Since the radio surveys that we use have different sensitivities, here we only considered the subsample of sources included in the NVSS survey with a radio flux density $\ge 3.5$ mJy.

From Fig.~\ref{lognsrr} we see that the surface density of the 3HSP subsample is approximately a factor of ten higher than that of the Sedentary Survey. This large difference is expected since the latter includes only extreme sources (its \nupeak\, distribution peaks at $\log \nu_{\rm peak} \sim$ 16.8,   compared to $\log \nu_{\rm peak} \sim$ 15.5 for the 3HSP sample). 
Similarly, for the case of DXRBS all BL Lacs outnumber the 3HSPs in every flux bin. 
For 3HSP subsamples, the higher the \nupeak\ cut, the lower the density, and the Sedentary number density are consistent with 3HSP subsample with a \nupeak\ cut at $10^{17.2}$~Hz. 
The DXRBS HBLs are also in very good agreement with the 3HSP number counts in the region of overlap, which shows that our selection criteria are robust.

Apart from the different normalisation, the $\log$N-$\log$S of every sample or subsample shows a similar trend deviating from the Euclidean slope at radio flux densities lower than $\approx 25$~mJy. 
The number densities for different BL Lac groups and different \nupeak\ HBL are almost parallel to each other, implying that the ratio of high \nupeak\ BL Lacs to low \nupeak\ BL Lacs remains the same regardless of the radio flux density. 
It is consistent with Figure~\ref{mjynu} which shows that the average radio flux density does not depend on synchrotron peak frequency and it seems that there is no preference for high \nupeak\ sources with faint flux and vice versa. 
This indeed deviates from the prediction of a `blazar sequence' scenario. 
A clear trend can be seen going from the Sedentary Survey of extreme HSPs to the 3HSP sample, to the entire population of BL Lacs as estimated in the DXRBS survey, with an increase in number by approximately a factor 10 at every step. 

The $\log$N-$\log$S indicates that both samples are complete at the bright end, but significantly deviate from the Euclidean slope, $-1.5$ (dashed line) at the faint end.
The Sedentary is a complete sample in the radio band, and the flattening at faint fluxes is due to its cosmological negative evolution, meaning that there are fewer faint radio sources in the young universe.
The 3HSP flattening, however, appears to be somewhat stronger than that of the Sedentary Survey, likely because of the onset of some degree of incompleteness at low radio flux densities in addition to the evolutionary effects discussed in the Sedentary Survey paper \citep{Piranomonte2007}.

The faintest bins of the 3HSP-NVSS-RASS subsample have a surface density $\approx 0.06~{\rm deg}^{-2}$, indicating that the maximum surface density corresponds to a total of $\sim 2,400$ HSP blazars over the whole sky. 
Given that this number refers only to sources with 1.4 GHz flux density $\ge 3.5$ mJy, and because of the incompleteness discussed above, this can be considered a robust lower limit.

Although the full 3HSP catalogue is not a complete sample in the radio band, we can build a complete subsample from it if we apply specific cuts. For instance, by using a radio and X-ray flux-limited sample and selecting only sources with \nupeak\ $\gtrsim 10^{16}$~Hz, we built a complete HSP subsample.
The subsample with limits in radio flux density, X-ray flux, and \nupeak\ value, comprising 1096 sources, can be considered highly complete and therefore useful for statistical purposes


\section{3HSP blazars as VHE source candidates}
\label{extreme}
High synchrotron peaked blazars are known to emit in the very high-energy (VHE) \gr\ band. 
By definition, the synchrotron \nupeak\ of HSPs are the highest among blazars, sometimes reaching and exceeding $10^{18}$~Hz.
Consequently, the peak frequency of the second hump in the SED of these extreme sources goes close to and sometimes well into the TeV band.
Synchrotron \nupeak\, values $\gsim 10^{18}$~Hz imply that the electrons responsible for the radiation must be accelerated to extremely high energies \citep{Rybicki1986,Costamante2001}. 

Considering a simple SSC model where $\nu_{\rm peak}=3.2 \times 10^6  \gamma^{2}_{\rm peak} B \delta $ \citep[e.g.][]{Giommi2012}, assuming $B=0.1$ gauss and a Doppler factor $\delta  =10$, HSPs characterised by $\nu_{\rm peak}$ ranging between $ 10^{15}$ and $\gsim  10^{18}$~Hz require $\gamma_{\rm peak}$ to be in the range 2$\times 10^4 - \gtrsim 10^6$, where $\gamma_{\rm peak}$ corresponds to the Lorentz factor of relativistic electrons emitting at the synchrotron peak frequency.  

Sources with \nupeak~$> 10^{17}$~Hz are often called extreme blazars and are particularly relevant for high-energy astrophysics since they may be the counterparts of VHE \gr\, sources, high-energy astrophysical neutrinos, and ultra high-energy cosmic rays (UHECR). 
Table~\ref{extremelist} gives the list of all objects in the 3HSP 
catalogue with  rest-frame \nupeak~$\ge 10^{17}$~Hz, and with  
\nupeak~$\ge 10^{17}$~Hz and no redshift, which implies that their 
rest-frame \nupeak\ is at least $10^{17}$~Hz. In total, there are 
384 extreme HSPs in our catalogue, a much larger number than in any 
previous catalogue. Based on their radio number counts (Fig. 
\ref{lognsrr}) we expect an all-sky content of about 370 extreme blazars down to the 3.5 mJy flux limit for the 1.4 GHz channel. Given that the number counts are 
relatively flat, we estimate that the total number of extreme blazars in the {\it sky} is $\approx 400$.

One of the main reasons for assembling the 3HSP catalogue was the need to find sources that could be detected in \gr\ surveys and provide promising targets for VHE/TeV observations. 
About 50\% of the 3HSP sources already appear in one of the existing \gr\ catalogues, sources in the other half are still undetected,
but a large fraction of them are expected to be above the sensitivity limit of upcoming VHE telescopes like the Cherenkov Telescope Array (CTA).

An example of this is PGC 2402248, a blazar recently detected by the Major Atmospheric Gamma Imaging Cherenkov Telescope (MAGIC) collaboration \citep{Mirzoyan2018} that is in the 2WHSP catalogue and is listed in 3HSP as a source with extreme \nupeak$=10^{17.9}$ Hz. 
The 1BIGB catalogue \citep{Arsioli2017} also demonstrates the potential of HSP catalogues built on the basis of multi-wavelength data for the detection and identification of \gr\, and VHE sources, and for the selection of targets for TeV observations. 

A significant fraction of {\it Fermi}-LAT detected sources still do not have   an assigned counterpart. 
As blazars are the dominant population of extragalactic persistent \gr\ sources, we checked whether some of the objects in the 3HSP sample could be the counterparts of still non-associated {\it Fermi} sources. 
We found that many 3HSPs may be the counterparts of {\it Fermi} 3FHL and 3FGL sources that still do not have a counterpart  assigned in the current Fermi-LAT catalogues. 
In Table~\ref{noFermiassoc} we list 19 possible counterparts for these non-associated {\it Fermi} 3FHL or 3FGL detections. 
These 19 sources   have no association counterpart in the recent 4FGL catalogue either. 

We note that \citet[][hereafter K19]{Kaur2019} presented the 
results of an identification campaign of unassociated sources from the {\it Fermi} 3FHL catalogue.
In Table~\ref{noFermiassoc}, we indicated those already in K19, and there are ten sources that are not in their identification list. 
Among their identifications, only 10 out of 110 sources are not included in the 3HSP. 
We checked all of them, and they are either non-HSP blazars or too uncertain to be included in the 3HSP catalogue. 
The other 91 sources (i.e. 110 minus 10, which are not in the 3HSP, and 9 already in Table~\ref{noFermiassoc}) either have an association in the 4FGL catalogue or in the 3FHL catalogue.

\begin{figure} [h!]
\begin{center}
\includegraphics[width=0.98\linewidth]{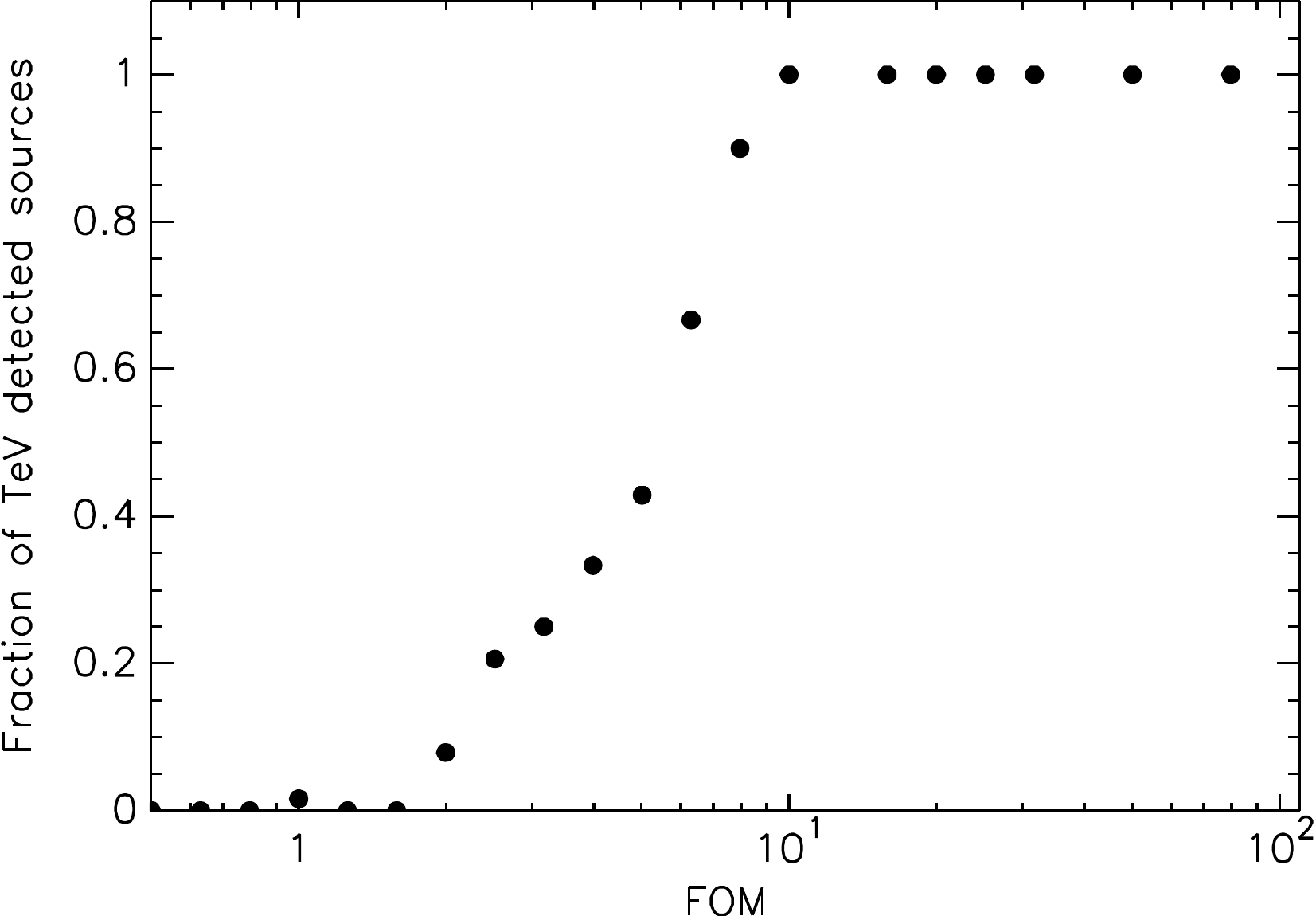}
\end{center}
\caption{Fraction of sources  detected in the TeV band for FOM values above the
one indicated.}
\label{fomtev}
\end{figure}

We applied the FOM value to estimate the potential detectability of HSPs by current and future VHE-TeV telescopes. 
FOM is defined in units of $2.5\times10^{-12}~{\rm erg}~{\rm s}^{-1}~{\rm cm}^{-2}$ (Log \nufnupeak~$=-11.6$), which is the peak flux of the faintest HSP blazar that is included in  TeVCat (see {Paper I}  for more details).
Figure~\ref{fomtev} indicates that the TeV 
detection rate increases with the FOM between 1 
and 10 and that the majority of cases with 
FOM$>$10 have already been reported in TeVCAT.
Sources with a higher FOM value (i.e. brighter flux) are expected to be more easily detectable by the current generation of VHE-TeV telescopes. 
We note that there are still 262 sources with FOM$\ge1$ that are not yet listed in TeVCAT, and 251 of them have already been detected by {\it Fermi} as reported in the  FGL or FHL catalogues.
Those sources should be of great interest to CTA, especially for the planning of observation campaigns.

\section{Conclusions}
We presented 3HSP, the largest compilation of high-synchrotron peaked blazars. 3HSP is an evolution of the 2WHSP \citep{Chang2017} and 1WHSP catalogues \citep{Arsioli2015}, which were assembled starting from Wise IR data, but were largely incomplete in the radio, X-ray, and \gr\ band.
The 3HSP catalogue contains 2013 HSPs and HSP-candidates, with 1618 of them also in the 2WHSP catalogue and 1007 having {\it Fermi} \gr\ counterparts. Only 657 3HSP sources are in the 5BZCat\citep{Massaro2015}, implying that the number of known HSP blazars has tripled compared to 2015 when BZCAT was the most complete list of blazars available. 
Another distinctive aspect of the 3HSP catalogue is that it provides redshift estimates for 88\% of the sources, a much higher percentage than in any previous catalogue. 


Providing robust candidates for high-energy, VHE/TeV, and \gr\ observations was one of the main motivations for building the 3HSP catalogue. 
Previous versions of the catalogue have already been used as a seed for VHE or TeV observations, and several new detections from HE or VHE has been secured based on the positions of 2WHSP sources. 
For example, a new VHE counterpart, PGC 2402248 (2WHSP J073326.7+515355), has recently been detected by MAGIC  \citep{Mirzoyan2018}, showing that the 3HSP can contribute potential VHE candidates for future surveys. 
The 1BIGB catalogue \citep{Arsioli2017} is another successful example that demonstrated the presence of new \gr\ sources found based on 2WHSP sources. 
Therefore, with multi-frequency-based catalogues the search for new VHE sources might become more efficient as 3HSP could point out the possible location of relevant VHE sources that were not known before. 
We note that 26 3HSP sources have been proposed as counterparts of non-associated {\it Fermi} sources (see Table~\ref{noFermiassoc}). 

We also presented several tests to investigate the completeness and the general properties of the catalogue. 
Our results suggest that the radio and X-ray selected 3HSP subsample is complete for \nupeak\ $> 10^{16}$~Hz. 
This complete and large (1096 sources) subsample is suitable for the detailed investigation of intrinsic statistical properties associated with HSP and extreme blazars. This will be presented in future publications. 

We note that after this paper was completed \cite{Paliya2019}
  applied a new $\gamma$-ray data analysis technique to 337 3HSP sources with \nupeak\ $> 10^{17}$ Hz. They detected 165 objects and report a cumulative signal at $> 32\sigma$ confidence for the remaining 172 {\it Fermi} $\gamma$-ray undetected 3HSP sources. 
Their average spectral slope is very flat ($\Gamma < 2$). 
This justifies the usefulness of the 3HSP catalogue in finding new VHE $\gamma$-ray sources.

\begin{acknowledgements}
YLC is supported by the Government of the Republic of China (Taiwan), BA is supported by S\~ao Paulo Research Foundation (FAPESP) with grant n. 2017/00517-4. PP thanks the SSDC for the hospitality and partial financial support of his visits. CHB would like to thank the Brazilian government and the CAPES foundation for supporting this work under the project BEX 15113-13-2. 
PG acknowledges the support of the Technische Universit\"at M\"unchen - Institute for Advanced Study, funded by the German Excellence Initiative (and the European Union Seventh Framework Programme under grant agreement no. 291763)
This work was supported by the SSDC, Agenzia Spaziale Italiana Science Data Center; and University Sapienza of Rome, Department of Physics. We made use of archival data and bibliographic information obtained from the NASA/IPAC Extragalactic Database (NED), data and software facilities from the SSDC managed by the Italian Space Agency (ASI).
\end{acknowledgements}

\bibliographystyle{aa}
\bibliography{3HSP.bib}

\begin{appendix}
\section{Association of no-radio sources around {\it Fermi} detections}
There are several sources within the 3HSP catalogue without radio detections. They were found by directly examining the area around {\it Fermi} detections reported in the 3FHL catalogue. 
Section~\ref{sec:building}  shows that checking \gr\ detected regions is a powerful way to search for blazar candidates. 
Here we show an example to illustrate how we estimated the synchrotron \nupeak\ and why these sources are robust HSP-candidates. 

\begin{figure} [h!]
\begin{center}
\includegraphics[width=1.0\linewidth]{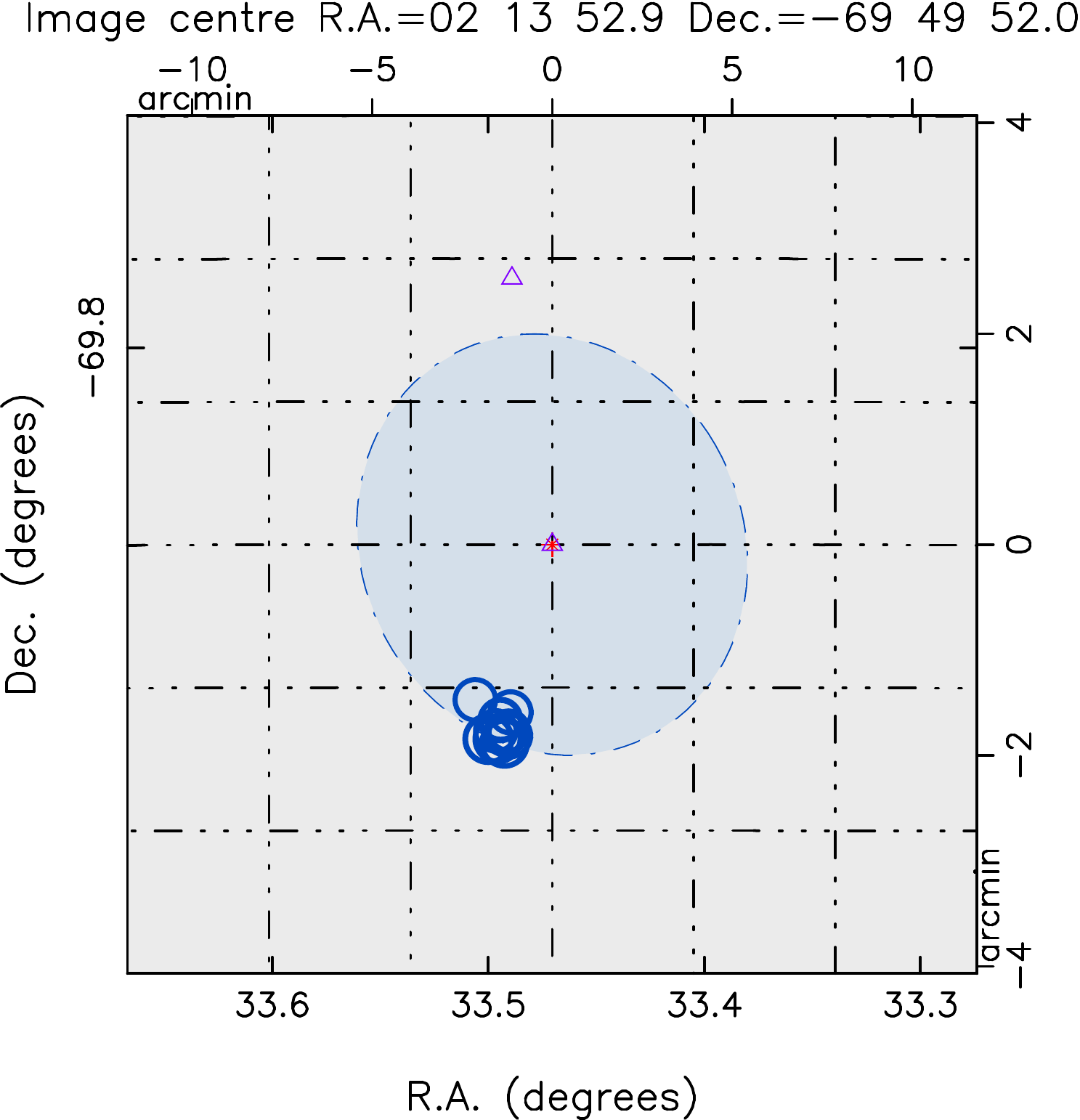}
\end{center}
\caption{Region around 4FGL J0213.8-6949. Purple triangles represent \gr\ detections (3FHL, 4FGL, and 9yr MST) and blue circles indicate X-ray detections. 6dF J0213586-695137 is the only X-ray source inside the Fermi detected region.}
\label{candidatemap}
\end{figure}


First, we discuss the case of 6dF J0213586-695137 (see section 2 and figure 2). 
We carefully checked the error region of the Fermi source detected as part of the 3FHL, 8-Year catalogue (4FGL) and 9yr MST \citep{Campana2018} lists and we confirm that 6dF J0213586-695137 is the only plausible counterpart (See Figure~\ref{candidatemap}). 
No {\it Fermi} blazars  emit in the radio or 
X-rays, so finding a {\it Fermi} counterpart can 
begin by checking radio sources and X-ray sources around it. 
Figure~\ref{candidatemap} shows the positions
of the $\gamma$-ray detections (from 3FGL, 3FHL, 
and 4FGL, with the error circle referring to the 
last: purple triangles) and of the X-ray detections of 6dfJ021358.6-695137 (from {\it XMM-Newton} slew and XRT: blue circles). The corresponding UV (GALEX), optical (GAIA), and IR (AllWISE) counterparts are all within $\approx 5-10$ arcsec from the X-ray positions. Based on the available information 
6dfJ021358.6-695137 is the only source  within the \gr\ error ellipse with a blazar-like optical/UV/X-ray SED consistent with the \gr\ data, and therefore the very likely counterpart of the {\it Fermi} \gr\ detection. 

\begin{figure} [h!]
\begin{center}
\includegraphics[width=1.0\linewidth]{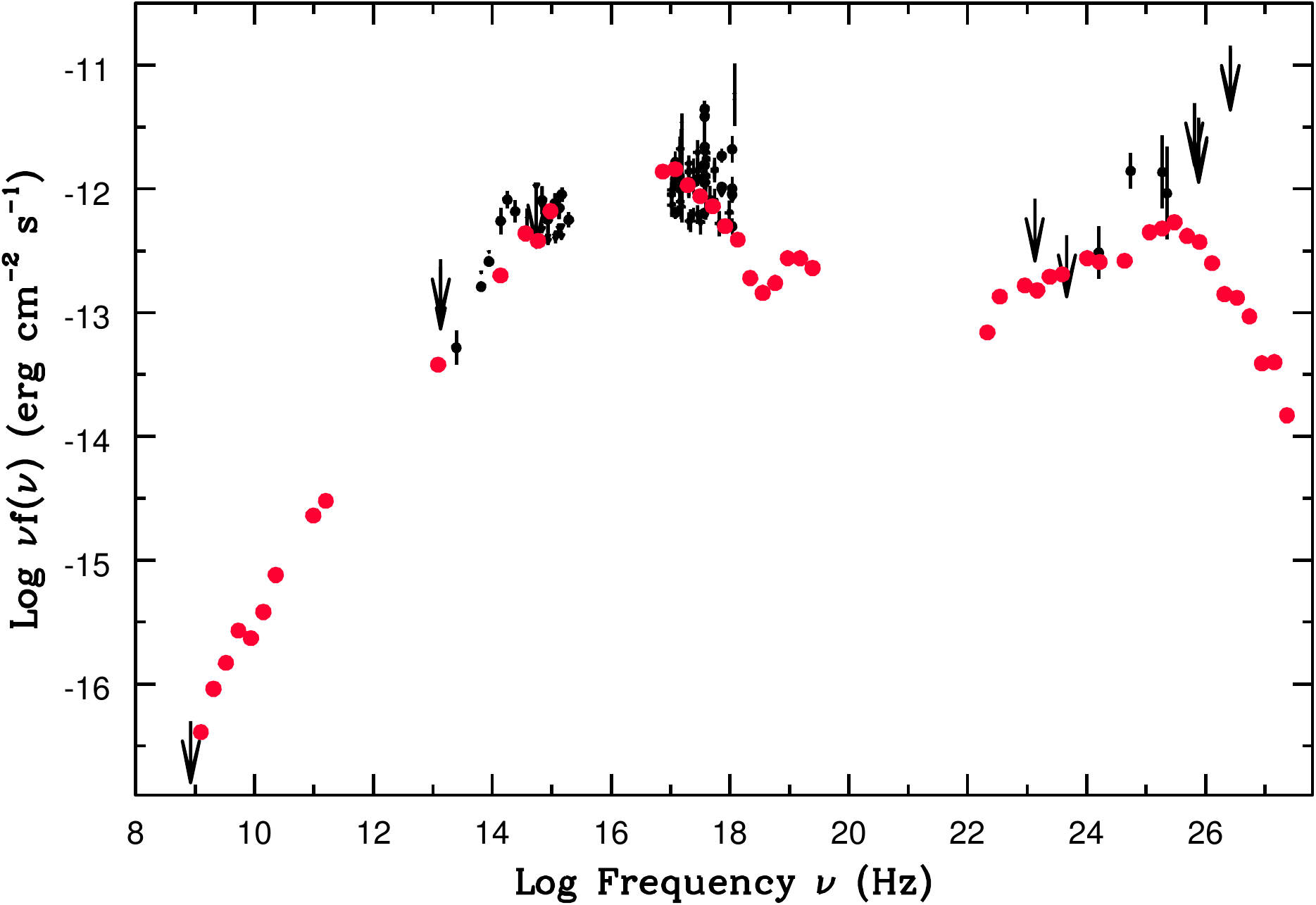}
\end{center}
\caption{Spectral energy distribution of 6dfJ021358.6-695137. The red filled circles represent the average SED of MKN421 scaled down to the flux level of 6dfJ021358.6-695137.}
\label{sed6df}
\end{figure}

Figure~\ref{sed6df} is the SED of 6dfJ021358.6-695137. 
The SED is fully consistent with that of an HSP blazar, such as a variable and flat X-ray spectrum and flat gamma-ray spectrum with Compton dominance $<~1$.
If the X-ray flux were to be attributed to a nearby AGN, then, based on the observed distribution of optical-UV/X-ray flux ratio ($\alpha_{\rm oX}$) of radio quiet AGN, this hypothetical object should be about 50 times brighter than 6dfJ021358.6-695137 in the optical and UV bands.
No such object is present in (or near) the X-ray error circles, so we can exclude that the X-ray source is a background AGN. 

We re-scaled the SED of Mrk 421, and plot the re-scaled fluxes of Mrk 421 in  figure~\ref{sed6df} (brown points), suggesting that the shape of the two SEDs (6dfJ021358.6-695137 and Mrk 421) are very similar. 
The extrapolation of the UV emission in the SED closely matches  the X-ray
spectrum.
Moreover, we do not claim that this is a radio quiet HSP since the radio upper limit in its SED is consistent with the flux predicted by the SED of MKN421 (rescaled to the flux of this source, and shown as a brown dotted line).  
The lack of radio detection in the 3HSP source is due to insufficient sensitivity of the SUMSS survey rather than to an usually low radio emission in the object. 

In conclusion, 6dfJ021358.6-695137 
(3HSPJ021358.6-695137) is the counterpart of the 
Fermi source and is clearly an extreme HSP blazar. 
This source is not found by chance or just because it is inside the {\it Fermi} detected region. 

There are only 24 sources in the 3HSP catalogue that do not have a radio counterpart, and we double checked   these source one by one again. 
We list all of them in Table 4 now. 
Among these sources, there are only three sources that do not have a UV counterpart yet. 
All three of these three sources have well-described X-ray spectral data, and we could fit the synchrotron peak from it. 
The other `no-radio' sources all have  IR, UV, optical, and X-ray counterparts, and we could clearly tell their peak frequency.
After checking all of them, we suggest that these sources are plausible HSPs, even though they have no radio counterparts. 

\begin{figure} [h!]
\begin{center}
\includegraphics[width=1.0\linewidth]{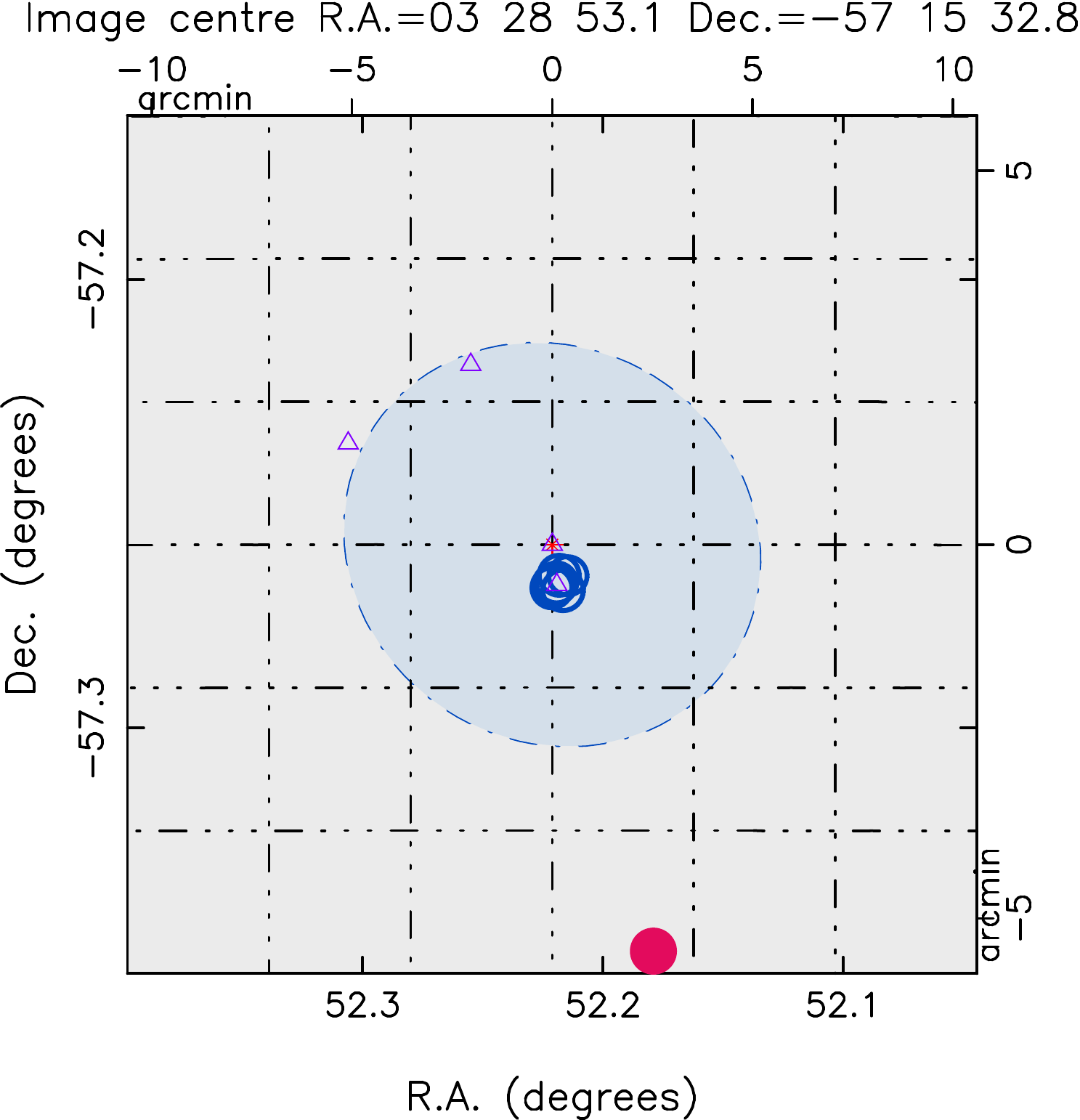}
\end{center}
\caption{Region around 4FGL J0328.8-5715. Purple triangles represent \gr\ detections (3FHL, 4FGL, and 1BIGB), blue open circles indicate  X-ray detections, while the red filled circle denotes an
unrelated radio source outside the \gr\ error circle. 3HSPJ032852.6-571605 is the only X-ray source inside the Fermi region. There are three \gr\ detections around this 3HSP sources, and 3HSPJ032852.6-571605 is the only plausible counterpart for them.}
\label{candidatemap2}
\end{figure}

\begin{figure} [h!]
\begin{center}
\includegraphics[width=1.0\linewidth]{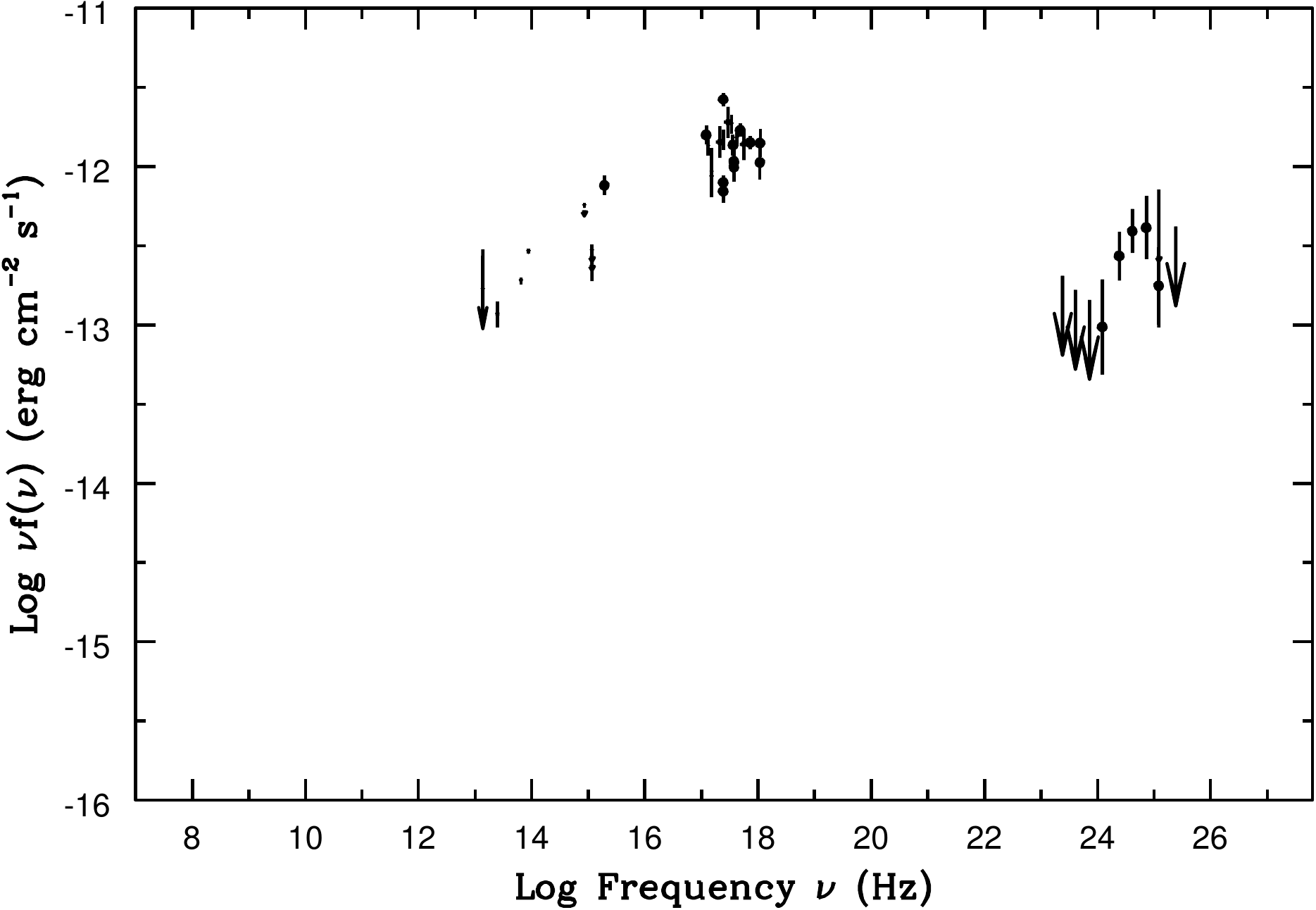}
\end{center}
\caption{Spectral energy distribution of 3HSPJ032852.6-571605. The Fermi spectrum is fully consistent (in intensity and slope) with the synchrotron part of the SED. The X-ray and the UV data imply a \nupeak\ $\approx 10^{17.5}$~Hz. }
\label{sednoradio2}
\end{figure}

Here we show another example, namely 3HSPJ032852.6-571605 (Figure~\ref{candidatemap2} to \ref{sednoradio2}). 





We conclude that those sources without a radio counterpart in our catalogues and inside {\it Fermi}-LAT error ellipses are genuine HSP-candidates. 
We did not select them  because they are inside the Fermi error circles or with X-ray detections, and every source in the 3HSPs was carefully examined. 
Keeping the focus on the aim of this catalogue, which is to provide seed sources for VHE observations, we include all promising HSP and HSP-candidates based on the currently available information. 
Thus, we could not exclude those sources.

In addition, these sources were not found by chance. 
According to the X-ray logN-logS, there might be a surface density of 0.06-0.07 deg$^{-2}$ HSPs. 
As the logN-logS suggests, we might find by chance one HSP within 14-15 square degrees, which is much larger than the area covered by the Fermi error circle. 
The largest position error area in the Fermi catalogues is around 1 degree.
In summary, the reported HSPs are reliable counterparts for the gamma-ray signatures, in agreement with their gamma-ray error circles, as reported by the Fermi team. This is so for all the gamma-ray counterparts reported in this work.  

\end{appendix}

\begin{center}
\begin{longtable}{>{\centering\arraybackslash}p{4.cm}>{\centering\arraybackslash}p{0.8cm}>{\centering\arraybackslash}p{1.2cm}>{\centering\arraybackslash}p{1.5cm}>{\centering\arraybackslash}p{0.8cm}>{\centering\arraybackslash}p{1.cm}>{\centering\arraybackslash}p{0.8cm}>{\centering\arraybackslash}p{3.3cm}>{\centering\arraybackslash}p{1.1cm}}
\caption{Sample of the 3HSP catalogue showing only the first 65 sources. Column 2 is the source flag:  1 is for uncertain sources,  2  for dubious photometric sources,  3  for relatively low peak flux--radio flux ratio sources, and  4  for ambiguous origins of IR and optical data (either from synchrotron or from host galaxy). Columns 5 and   7 are flags for synchrotron peak frequency and redshift, respectively. 
For each column and flag, see text for details:  1 is for `firm estimate',   2 for `uncertain', and   3 for `lower limits'. Flags 4 and  5 are for photo-z. Column 8 lists the \gr\ counterpart for each source. For those beginning with J instead of a catalogue name (such as 3FGL, 4FGL) the \gr\ detections are from the 1BIGB catalogue \citep{Arsioli2017}. Column 9 is the FOM,  defined in units of $2.5\times10^{-12}~{\rm erg}~{\rm s}^{-1}~{\rm cm}^{-2}$ (Log \nufnupeak~$=-11.6$), which is the peak flux of the faintest HSP blazar that has been detected in the TeVCat (see 1WHSP paper for more details). The full table of the 3HSP catalogue is available at http://www.asdc.asi.it/3hsp/}\\
\hline\hline
Source&flag&2WHSP&Log \nupeak&flag&Z&flag&\gr\ counterpart&FOM\\
\hline
\endfirsthead
\hline\hline
Source&flag&2WHSP&Log \nupeak&flag&Z&flag&\gr\ counterpart&FOM\\
\hline
\endhead
\hline
\endfoot 
3HSPJ000116.4+293534 & 1 & yes & 16.0 & 2 & 0.58  & 5  & --               & 0.05  \\   
3HSPJ000117.1$-$315043 & 1 & yes & 15.5 & 2 & 0.45  & 4  & --               & 0.2   \\ 
3HSPJ000132.7$-$415525 & 4 & yes & 15.8 & 1 & --    & -- & 4FGLJ0001.6$-$4156 & 0.79  \\ 
3HSPJ000158.1$-$115047 & 1 & yes & 16.2 & 1 & 0.53  & 5  & --               & 0.13  \\ 
3HSPJ000215.2$-$672653 & 0 & yes & 15.8 & 1 & 0.52  & 3  & 4FGLJ0002.1$-$6728 & 0.63  \\ 
3HSPJ000236.1$-$081532 & 1 & yes & 16.0 & 1 & 0.39  & 5  & --               & 0.2   \\ 
3HSPJ000319.6$-$524727 & 0 & yes & 16.9 & 1 & 0.37  & 5  & 4FGLJ0003.1$-$5248 & 0.5   \\ 
3HSPJ000513.7$-$261438 & 0 & yes & 15.4 & 1 & 0.32  & 1  & --               & 0.1   \\
\end{longtable}
\end{center}

\begin{center}
\begin{longtable}{>{\centering\arraybackslash}p{4.cm}>{\raggedleft\arraybackslash}p{1.cm}>{\centering\arraybackslash}p{1.1cm}>{\centering\arraybackslash}p{0.5cm}>{\centering\arraybackslash}p{1.cm}>{\centering\arraybackslash}p{1.2cm}>{\centering\arraybackslash}p{1.2cm}>{\centering\arraybackslash}p{2.9cm}>{\centering\arraybackslash}p{2.5cm}}
\caption{Extreme synchrotron peak sources. Column 4 is the flag for redshift:  1 is for `firm estimate',   2 for `uncertain', and  3 for `lower limit'. Flags 4 and  5 represent photo-z. Column 5 is the FOM,  defined in units of $2.5\times10^{-12}~{\rm erg}~{\rm s}^{-1}~{\rm cm}^{-2}$ (Log \nufnupeak~$=-11.6$), which is the peak flux of the faintest HSP blazar  detected in the TeVCat   to date (see 1WHSP paper for more details). Column 8 lists the \gr\ counterpart for each source. For those beginning with J instead of a catalogue name (such as 3FGL, FY8Y), the \gr\ detections are from the 1BIGB catalogue \citep{Arsioli2017}.}\\
\hline\hline
Source&$\log$~\nupeak&z&flag&FOM&BZCat&2WHSP&Gamma&TeVCat\\
&Hz& & & & & & & \\
\hline
\endfirsthead
\caption[]{continued.}\\
\hline\hline
Source&$\log$~\nupeak&z&flag&FOM&BZCat&2WHSP&Gamma&TeVCat\\
&Hz& & & & & & & \\
\hline
\endhead
\hline
\endfoot 
3HSPJ000835.4$-$233927 & 17.0    & 0.147 & 1  & 0.79  & 5BZB & yes & 4FGLJ0008.4$-$2339 & --             \\
3HSPJ001328.1+094930 & 17.3    & 0.29  & 5  & 0.63  & --   & yes & J001328.8+094929 & --             \\
3HSPJ001356.0$-$185406 & 17.4    & 0.094 & 1  & 2.0   & 5BZG & yes & 4FGLJ0013.9$-$1854 & SHBL           \\
3HSPJ001411.5$-$502234 & 17.5    & 0.01  & 2  & 1.0   & 5BZB & yes & 4FGLJ0014.1$-$5022 & --             \\
3HSPJ001827.8+294730 & 17.1    & 0.1   & 2  & 1.58  & 5BZB & yes & 4FGLJ0018.4+2946 & --             \\
3HSPJ003322.5$-$203908 & 17.9    & 0.073 & 1  & 0.5   & --   & yes & --               & --             \\
3HSPJ003552.6+595004 & 18.2    & 0.086 & 2  & 19.95 & 5BZB & --  & 4FGLJ0035.9+5950 & 1ES0033+595    \\
3HSPJ004013.1+405004 & $>$17.5 & 0.24  & 5  & 1.26  & 5BZU & yes & 4FGLJ0040.3+4050 & --             \\
\end{longtable}
\end{center}

\begin{table*} [h!]
\caption{Fermi sources reported as unassociated, but with a counterpart in the 3HSP catalogue.}
\begin{center}
\begin{tabular}{ccrccc}
\hline\hline
Source&R.A.&Dec.&4FGL&3FHL& In Kaur et al. 2019\\
\hline
3HSPJ015624.5$-$242003 & 29.10228  & $-$24.33438 & J0156.5$-$2420 & J0156.2$-$2419& yes\\
3HSPJ025111.5$-$183112 & 42.79803  & $-$18.52021 & J0251.1$-$1830 & J0251.2$-$1830& yes\\
3HSPJ025857.5+055243 & 44.73981  & 5.87889   & J0259.0+0552 & J0258.9+0554&no\\
3HSPJ043837.1$-$732921 & 69.65448  & $-$73.48933 & J0438.0$-$7329 & J0438.0$-$7328& yes\\
3HSPJ072547.9$-$054832 & 111.4495  & $-$5.809    & J0725.7$-$0549 & J0725.7$-$0548& yes\\
3HSPJ074710.0$-$073724 & 116.79181 & $-$7.62359  & J0747.2$-$0736 & ---       &no\\      
3HSPJ091926.2$-$220042 & 139.85933 & $-$22.01187 & J0919.5$-$2200$^{*}$ & ---       &no\\      
3HSPJ101620.7$-$424722 & 154.08614 & $-$42.78961 & J1016.1$-$4247 & J1016.2$-$4245& yes\\
3HSPJ114600.8$-$063854 & 176.50357 & $-$6.64859  & J1146.0$-$0638 & J1145.9$-$0637& yes\\
3HSPJ115514.9$-$111122 & 178.81193 & $-$11.18959 & J1155.2$-$1111 & ---             &no \\
3HSPJ122014.5$-$245948 & 185.06058 & $-$24.99685 & J1220.1$-$2458 & J1220.1$-$2459& yes\\
3HSPJ122327.5+082030 & 185.8646  & 8.34183   & J1223.5+0818 & ---             &no\\
3HSPJ141046.0+740511 & 212.69171 & 74.08647  & J1410.7+7405 & J1410.8+7406&no\\
3HSPJ194333.8$-$053353 & 295.89084 & $-$5.56492  & J1944.0$-$0535$^{*}$ & ---             &no\\
3HSPJ210415.9+211808 & 316.06635 & 21.30228  & J2104.4+2116 & J2104.5+2117&no\\
3HSPJ211522.0+121802 & 318.84169 & 12.30074  & J2115.2+1218 & J2115.2+1218& yes\\
3HSPJ214239.8$-$202819 & 325.66579 & $-$20.472   & J2142.5$-$2029 & J2142.5$-$2029&no\\
3HSPJ214429.6$-$563849 & 326.12322 & $-$56.64697 & J2144.6$-$5640$^{*}$ & ---             &no\\
3HSPJ230848.7+542611 & 347.20309 & 54.43645  & J2309.0+5428$^{*}$ & J2308.8+5424& yes\\
\hline\hline
\end{tabular}
\label{noFermiassoc}
\end{center}
$^{*}$sources not in 4FGL, but in 3FGL.
\end{table*}

\begin{table*} [h!]
\caption{Sources without radio counterparts.}
\begin{center}
\begin{tabular}{ccrccc}
\hline\hline
Source&$\nu$ peak.&z&z flag&Gamma& FOM\\
\hline
3HSPJ015624.5$-$242003 & 15.7 & --   & -- & 4FGLJ0156.3$-$2420 & 0.5  \\  
3HSPJ020020.9$-$410935 & 15.5 & 0.5  & 2  & 4FGLJ0200.3$-$4109 & 0.32 \\
3HSPJ023905.5+132721 & 15.0 & --   & -- & --               & 0.2  \\
3HSPJ032852.7$-$571605 & 17.3 & 0.48 & 3  & 4FGLJ0328.8$-$5715 & 0.5  \\
3HSPJ040111.0$-$535458 & 16.8 & 0.59 & 5  & 4FGLJ0401.0$-$5353 & 0.25 \\
3HSPJ042011.0$-$601505 & 16.2 & 0.33 & 5  & 4FGLJ0420.3$-$6016 & 0.32 \\
3HSPJ045937.0$-$541707 & 17.2 & 0.5  & 5  & 4FGLJ0459.7$-$5413 & 0.16 \\
3HSPJ054106.9$-$485410 & 16.6 & 0.6  & 5  & 4FGLJ0541.1$-$4854 & 0.13 \\
3HSPJ073927.4$-$672136 & 17.0 & 0.53 & 5  & 4FGLJ0739.8$-$6722 & 0.5  \\
3HSPJ081003.3$-$752723 & 15.1 & 0.47 & 3  & --               & 0.79 \\
3HSPJ101620.7$-$424722 & 15.5 & 0.25 & 5  & 4FGLJ1016.1$-$4247 & 0.4  \\
3HSPJ102432.4$-$454426 & 17.0 & 0.37 & 5  & 4FGLJ1024.5$-$4543 & 0.32 \\
3HSPJ103332.2$-$503528 & 15.7 & 0.24 & 5  & 4FGLJ1033.5$-$5035 & 1.0  \\
3HSPJ103438.5$-$464403 & 17.1 & 0.33 & 5  & 4FGLJ1034.7$-$4645 & 0.5  \\
3HSPJ111715.2$-$533813 & 15.9 & --   & -- & 4FGLJ1117.2$-$5337 & 0.4  \\
3HSPJ124021.0$-$714857 & 17.4 & 0.21 & 5  & 4FGLJ1240.4$-$7148 & 1.58 \\
3HSPJ141046.0+740511 & 15.1 & --   & -- & 4FGLJ1410.7+7405 & 0.16 \\
3HSPJ151444.0$-$772254 & 16.3 & --   & -- & 4FGLJ1514.4$-$7719 & 0.5  \\
3HSPJ154439.4$-$112804 & 18.4 & --   & -- & 4FGLJ1544.5$-$1126 & 0.63 \\
3HSPJ154458.9$-$664146 & 17.3 & 0.23 & 5  & 4FGLJ1545.0$-$6642 & 1.26 \\
3HSPJ163146.7+414632 & 15.7 & 0.58 & 5  & 4FGLJ1631.8+4144 & 0.1  \\
3HSPJ184229.8$-$584157 & 17.1 & 0.33 & 5  & 4FGLJ1842.4$-$5840 & 1.58 \\
3HSPJ215936.1$-$461953 & 15.9 & 0.4  & 5  & 4FGLJ2159.6$-$4620 & 0.25 \\
3HSPJ224123.5+294247 & 17.0 & 0.48 & 5  & 4FGLJ2241.3+2943 & 0.4  \\
\hline\hline
\end{tabular}
\label{noradio}
\end{center}
\end{table*}

\end{document}